\documentclass[twocolumn,prl,aps,superscriptaddress,floatfix,showpacs]{revtex4-1}

\usepackage{multirow,amssymb,amsbsy,amsmath}
\usepackage{graphicx}
\usepackage{verbatim}
\usepackage{booktabs}
\usepackage{dcolumn}
\makeatletter

\usepackage{color}
\usepackage{xcolor}
\usepackage[normalem]{ulem}
\begin{document}

\title{Reconfigurable optical implementation of quantum complex networks
}
\author{Johannes Nokkala} 
\affiliation{Turku Centre for Quantum Physics, Department of Physics and Astronomy, University of Turku, FI-20014, Turun
Yliopisto, Finland}
\author{Francesco Arzani}
\affiliation{Laboratoire Kastler Brossel, Sorbonne Universit\'e, CNRS, ENS-PSL Research University, Coll\`ege de France; 4 place Jussieu, F-75252 Paris, France}
\author{Fernando Galve}
\affiliation{IFISC (UIB-CSIC), Instituto de Fisica Interdisciplinar y Sistemas Complejos, UIB Campus, 07122 Palma de Mallorca, Spain.}
\author{Roberta Zambrini}
\affiliation{IFISC (UIB-CSIC), Instituto de Fisica Interdisciplinar y Sistemas Complejos, UIB Campus, 07122 Palma de Mallorca, Spain.}
\author{Sabrina Maniscalco} 
\affiliation{Turku Centre for Quantum Physics, Department of Physics and Astronomy, University of Turku, FI-20014, Turun
Yliopisto, Finland}
\author{Jyrki Piilo} 
\affiliation{Turku Centre for Quantum Physics, Department of Physics and Astronomy, University of Turku, FI-20014, Turun
Yliopisto, Finland}
\author{Nicolas Treps}
\affiliation{Laboratoire Kastler Brossel, Sorbonne Universit\'e, CNRS, ENS-PSL Research University, Coll\`ege de France; 4 place Jussieu, F-75252 Paris, France}
\author{Valentina Parigi}
\affiliation{Laboratoire Kastler Brossel, Sorbonne Universit\'e, CNRS, ENS-PSL Research University, Coll\`ege de France; 4 place Jussieu, F-75252 Paris, France}

\begin{abstract}
Network theory has played a dominant role in understanding the structure of complex systems and their dynamics. 
Recently, quantum complex networks, i.e.~collections of quantum systems arranged in a non-regular topology, have been theoretically explored  leading to significant progress in a multitude of diverse contexts including, e.g., quantum transport, open quantum systems, quantum communication, extreme violation of local realism, and quantum gravity theories. 
Despite important progress in several quantum platforms, the implementation of complex networks with arbitrary topology in quantum experiments is still a demanding task, especially if we require both a significant size of the network and the capability of generating arbitrary topology - from regular to any kind of non-trivial structure - in a single setup. Here we propose an all optical and reconfigurable implementation of quantum complex networks. The experimental proposal is based on optical frequency combs, parametric processes, pulse shaping and multimode measurements allowing the arbitrary control of the number of the nodes (optical modes) and topology of the links (interactions between the modes) within the network. Moreover, we also show how to simulate quantum dynamics within the network combined with the ability to address its individual nodes.
To demonstrate the versatility of these features, we discuss the implementation of two recently proposed probing techniques for quantum complex networks and structured environments.
\end{abstract}

%
%

%
%
%
\maketitle

During the last twenty years, network theory has experienced remarkable progress and revolutionised the research in diverse disciplines ranging, e.g., from technology to social sciences and biology~\cite{RMPBarabasi,Newmann,Newman2,Trabesinger}. The discovery of new types of networks, such as having scale-free \cite{Goh02} or small-world properties \cite{Amaral00}, and development of new tools, e.g. community detection \cite{Rosvall08}, has led to the invaluable role that network theory currently has in empirical studies of many real-world complex systems. \\
By character, the combination of network theory and its applications are multidisciplinary and during the recent years a new area applying network theory and complex networks to quantum physical systems has emerged~\cite{Bianconi2015,Biamonte17}. Furthermore, specific features of quantum networks with no classical equivalent have been reported~\cite{Acin07,Faccin14,Paparo13,Perseguers10}. \\
Indeed, due to their ubiquitous nature, complex networks have found applications also in diverse topics in theoretical physics. 
The properties and topology of the underlying complex network influence localization properties of coherent excitations~\cite{PhysRevLett.101.175702}, phase transition of light
\cite{PhysRevE.87.022104}, Bose-Einstein condensation of non-interacting bosons~\cite{0953-4075-34-23-314}, and quantum walks with the associated quantum transport~
\cite{MULKEN201137}.
With technological advances for quantum communication and development of quantum internet~\cite{Kimble08}, the future information networks have to 
handle not only their computational complexity but also their complex geometrical structure, and there are already results on how the network topology influences entanglement percolation~\cite{Acin07,PhysRevLett.103.240503}. 
 Moreover, complex graphs of higher dimensions are strongly connected with special quantum algorithms~\cite{Rossi14} and networks have also been used for studies in quantum gravity~\cite{Bianconi16}.
In general, earlier theoretical discussions on quantum complex networks include topics from the control, probing and engineering of the network
\cite{Kato14,Jurcevic14,Elliott16,Moore16,Gokler17}, to the generation and protection of quantum resources \cite{Yung05,Kostak07,Moore16,Manzano13}, as well as the study of collective phenomena \cite{Arena08,Dofler13}, like quantum synchronization \cite{Manzano13sy,Bellomo17}. \\
However, despite the theoretical progress, a large gap remains  hindering the practical developments: Is it actually possible to create and control quantum complex networks with arbitrary topology in a single setup? \\ We give a positive answer for this question by combining earlier theoretical work on complex networks of interacting harmonic oscillators~\cite{Manzano13sy,Nokkala16} with recent experimental advances in creating multipartite entangled states in a multimode optical system~\cite{Roslund14}. Interacting optical modes represent an interesting option because they are highly resilient to noise, higly controllable with classical instruments and efficiently detectable.  Our proposal is based on a compact optical setup including optical frequency combs and parametric processes which, along with mode selective and multimode homodyne measurements, allows for the implementation of networks with reconfigurable coupling and topology, with sufficient size and diversity to be relevant in the context of complex networks. Here, the modes represent the nodes in the network and the interactions between them the links. 
We want also to stress here that although the efficient implementation of quantum networks has been achieved in other platforms, e.g. superconducting qubit \cite{Sameti17}, atoms in optical lattices \cite{Robens15} and single photons \cite{Qiang16}, the topologies which have been explored are mainly regular ones (lattice, circulant graphs, triangular graphs) and they mainly concern logical encoding rather than physical interactions between the quantum systems. On the contrary, our platform allows for: the deterministic implementation of the network, as the mapping is based on continuous variables, and the implementation of arbitrary complex topology within the limits of the experimentally achievable size. Moreover, the reconfigurability, i.e. varying the topology for the network, does not require to change the optical setup. \\
 The quantum nature of the networks is provided by the ability to initialize the oscillators in quantum states and/or generate entanglement connections between nodes.  
Part of this strategy has already demonstrated the fabrication of multipartite entangled states~\cite{Gerke15} and cluster states ~\cite{Roslund14,Cai17}.  Here we address a more general scenario, with additional tools and a specific mapping, for the implementation of quantum complex networks. \\
Several features make the scheme truly appealing for the general study of quantum complex networks. 
We can control the number of nodes in the network and in principle any network -- whether complex or not -- can be created from a given set of nodes. Moreover, the system allows to simulate quantum dynamics within the network by mapping the dynamical results of Ref.~\cite{Nokkala16} for optimized experimental parameters of the optical multimode set-up. It is also important to note that each node of the network can be individually addressed which opens significant possibilities to probe the global properties of the network by detecting the local properties, as proposed in~\cite{galve2013,Nokkala16}.  The proposal also opens the possibility to design quantum simulators for continuous variable open quantum systems. This is important because open system dynamics and memory effects have been very actively studied  in the last ten years both theoretically~\cite{0034-4885-77-9-094001,RevModPhys.88.021002,RevModPhys.89.015001} and experimentally (see, e.g.,~\cite{Liu:2011aa}). On the other hand, memory effects have been predicted considering microscopic bath realization of coupled oscillators \cite{Vasile14} and can be realized with the present set-up. However, most of the experimental work in this context has dealt with conceptually rather simple single qubit (photon) open systems and no experiments on controlled continuous variable open system dynamics with memory effects exist yet, to the best of our knowledge. 
Our results also benefit research on complex quantum communication  and information networks ~\cite{Perseguers10,Paparo12}, and in energy transport and harvesting within biological systems
\cite{Walschaers16rev,Faccin14}.  In the following sections, we first describe the dynamics within the complex network of coupled harmonic oscillators and then show in detail how  this can be mapped  to the optical platform. We then continue by demonstrating the implementation of two probing techniques for complex networks,  and discuss the requirements of scalability and reconfigurability of the system.

\section{Network dynamics}

We consider bosonic  quantum complex networks composed by an ensemble of quantum harmonic oscillators, with frequencies $\omega_i$, linked by spring-like couplings according to a specific topology. This is  defined by the adjacency matrix $\mathbf{V}$, which contains the coupling terms $v_{ij}$ for any couple of nodes. Notice that since the interactions between the oscillators are symmetric, the adjacency matrix must be symmetric as well, meaning that the networks will be undirected.
\begin{figure}[ht]
  \includegraphics[width=\linewidth]{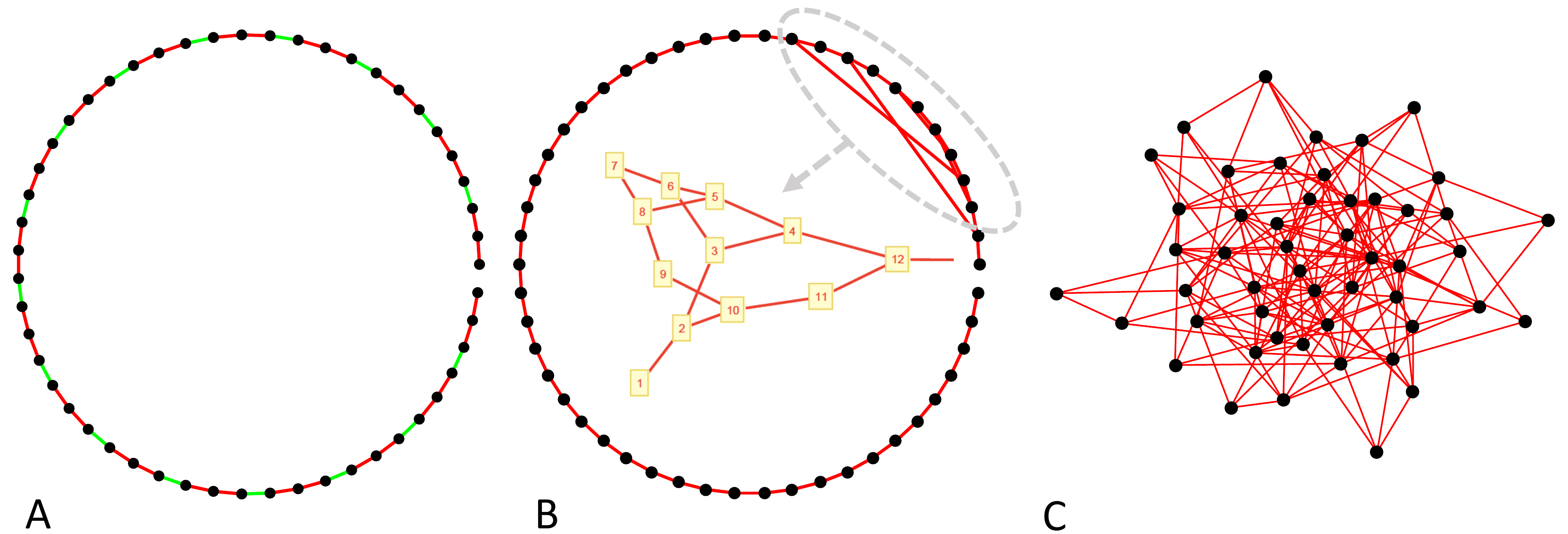}
  \caption{Graphical structure of the investigated networks. \textbf{A} is  a periodical chain with 51 nodes,  the coupling strength  $v_{ij}=0.1$,  is reduced to $v_{ij}=0.06$ every third link. \textbf{B} is a linear chain, with $v_{ij}=0.1$, where weak shortcuts with strength $v_{ij}/50$ are added.  In the center the shortcuts structure is shown. \textbf{C} is  an Erd\H{o}s-R\'{e}nyi random network  of 50 nodes with $v_{ij}=0.01$.}
  \label{fig:net}
\end{figure}
 This system can provide a quantum model of a heat bath sustained by an electromagnetic field, or it can be interpreted as a generic bosonic structure which can be exploited, e.g. for transporting  energy or information. 
The network Hamiltonian is $H_E=\mathbf{p}'^T\mathbf{p}'/2+\mathbf{q}'^T\mathbf{Aq}'$ where we have vectors of momentum and position operators $\mathbf{p}'^T=\{p_1',...,p_N'\}$, $\mathbf{q}'^T=\{q_1',...,q_N'\}$ for N harmonic oscillators. The matrix $\mathbf{A}$ has diagonal elements which can be written by introducing the effective frequencies $\tilde{\omega}_i$ as $\mathbf{A}_{ii}=\tilde{\omega}_i^2/2=\omega_i^2/2+\sum_j v_{ij}/2$, and off-diagonal elements $\mathbf{A}_{i\neq j}=-v_{ij}/2$.  Note that $\mathbf{A}$ can be expressed in terms of the adjacency matrix $\mathbf{V}$ of the network and a diagonal matrix  containing the effective frequencies $\mathbf{\Delta}_{\tilde{\omega}}=\textup{diag}\{\tilde{\omega}_1^2,...,\tilde{\omega}_N^2\}$, as $\mathbf{A}=\mathbf{\Delta}_{\tilde{\omega}}/2-\mathbf{V}/2$.

\begin{figure}
  \includegraphics[width=\linewidth]{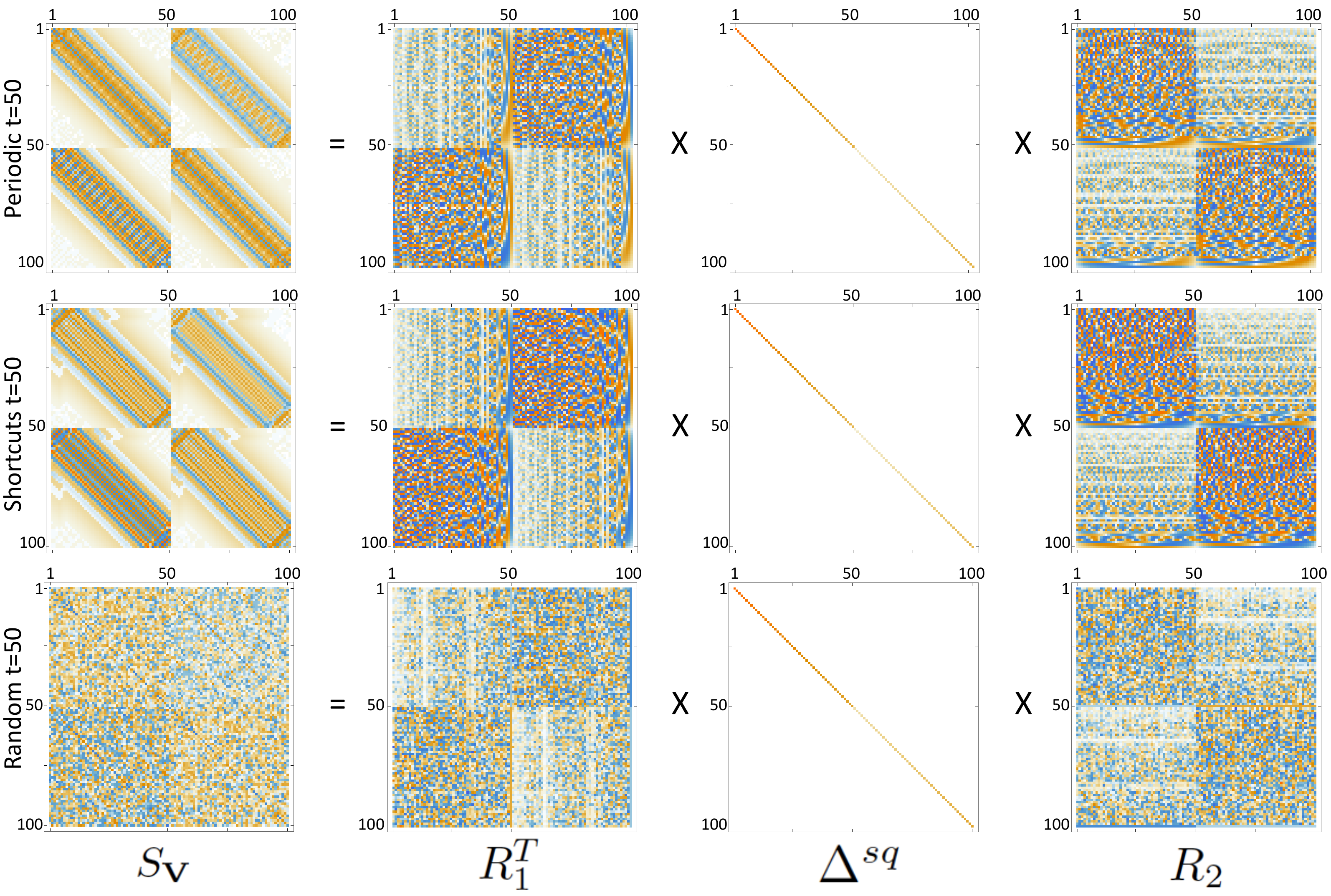}
  \caption{B-M decomposition for the three networks at t=50. The evolution $S_\mathbf{V}$ can be decomposed as the product of a squeezing operation $\Delta^{sq}$ in between two linear optics transformations $R_1^T$ and  $R_2$. }
  \label{fig:BM}
\end{figure}

As the bosonic oscillators are intended to describe electromagnetic fields we use a more common formalism in quantum optics by substituting  $\mathbf{q}'$ and $\mathbf{p}'$ with the renormalized quadratures $\mathbf{q}$ and  $\mathbf{p}$ by defining $ \mathbf{q}^T =\mathbf{q}'^T\sqrt{\mathbf{\Delta}_\omega}$ and $\mathbf{p}^T=\mathbf{p}'^T\sqrt{\mathbf{\Delta}_\omega^{-1}}$ with $\mathbf{\Delta}_\omega=\textup{diag}\{\omega_1,...,\omega_N\}$. So the network Hamiltonian is rewritten as
\begin{equation}\label{HE}
H_E=\dfrac{\mathbf{p}^T\mathbf{\Delta}_\omega\mathbf{p}}{2}+\mathbf{q}^T\sqrt{\mathbf{\Delta}_\omega^{-1}}\mathbf{A}\sqrt{\mathbf{\Delta}_\omega^{-1}}\mathbf{q}
\end{equation}
By  diagonalizing the symmetric matrix  $\mathbf{A}$ with an orthogonal matrix $\mathbf{K}$, such as $\mathbf{K}(\mathbf{\Delta}_\Omega/\sqrt{2})^2\mathbf{K}^T=\mathbf{A}$, where $\mathbf{\Delta}_\Omega$ is a diagonal matrix $=\textup{diag}\{\Omega_1,...,\Omega_N\}$, we can define decoupled oscillators, also named normal modes, with quadratures
\begin{align}
\label{QP}
\begin{split}
 \mathbf{Q}=\sqrt{\mathbf{\Delta}_\Omega}\mathbf{K}^T\sqrt{\mathbf{\Delta}_\omega^{-1}} \mathbf{q},
\\
 \mathbf{P}=\sqrt{\mathbf{\Delta}_\Omega^{-1}}\mathbf{K}^T\sqrt{\mathbf{\Delta}_\omega} \mathbf{p}.
\end{split}
\end{align}
Plugging Eq. (\ref{QP}) in Eq. (\ref{HE}) the network Hamiltonian becomes
\begin{equation}
H_E=(\mathbf{P}^T\mathbf{\Delta}_\Omega\mathbf{P}+\mathbf{Q}^T\mathbf{\Delta}_\Omega\mathbf{Q})/2=\sum_i^N\Omega_i(P_i^2+Q_i^2)/2.
\end{equation} 
From the free temporal evolution of the decoupled quadratures $\{\mathbf{Q},\mathbf{P}\}$ and by inverting the relations (\ref{QP}) we can recover the temporal evolution of the network oscillators: 
\begin{equation}\label{netoscEM}
\begin{pmatrix}
\mathbf{q}(t) \\
\mathbf{p}(t) \\
\end{pmatrix}  = 
\begin{pmatrix}
T_1 D^{\Omega}_{\cos}  & T_1 D^{\Omega}_{\sin}  \\
-T_2 D^{\Omega}_{\sin} & T_2 D^{\Omega}_{\cos} 
\end{pmatrix}
\begin{pmatrix}
\mathbf{Q}(0) \\
\mathbf{P}(0) \\
\end{pmatrix} 
\end{equation}
 and then
\begin{equation}\label{netosc}
\begin{pmatrix}
\mathbf{q}(t) \\
\mathbf{p}(t) \\
\end{pmatrix}  = 
\begin{pmatrix}
T_1 D^{\Omega}_{\cos} T_1^{-1} & T_1 D^{\Omega}_{\sin} T_2^{-1} \\
-T_2 D^{\Omega}_{\sin}T_1^{-1} & T_2 D^{\Omega}_{\cos} T_2^{-1}
\end{pmatrix}
\begin{pmatrix}
\mathbf{q}(0) \\
\mathbf{p}(0) \\
\end{pmatrix} 
\end{equation}
where $ T_1=\sqrt{\mathbf{\Delta}_\omega}\mathbf{K}\sqrt{\mathbf{\Delta}_\Omega^{-1}}$  and $ T_2=\sqrt{\mathbf{\Delta}_\omega^{-1}}\mathbf{K}\sqrt{\mathbf{\Delta}_\Omega}$ and we have introduced the diagonal matrices $D^{\Omega}_{\sin ii}=\sin(\Omega_i t)$ and $D^{\Omega}_{\cos ii}=\cos(\Omega_i t)$.  The equations can be given  in the compact form 
\begin{align}
\label{Xx}
\begin{split}
 \mathbf{x}(t)=S'_\mathbf{V} \mathbf{X}(0),
\\
 \mathbf{x}(t)=S_\mathbf{V} \mathbf{x}(0)
\end{split}
\end{align}
where   $\mathbf{X}^T=\{\mathbf{Q}^T,\mathbf{P}^T\}$ and  $\mathbf{x}^T=\{\mathbf{q}^T,\mathbf{p}^T\}$.
The optical implementation of the evolution described by (\ref{netoscEM}) or (\ref{netosc}) requires a multimode system, consisting of a number of modes equal to the number of nodes in the network. The optical modes need to be prepared in the initial state described  by the quadratures $\{\mathbf{Q(0)}^T,\mathbf{P(0)}^T\}$ in the normal modes picture or, equivalently, by  $\{\mathbf{q(0)}^T,\mathbf{p(0)}^T\}$ in the bare oscillators picture. Finally the implementation involves reconfigurable optical processes, able to drive the quadratures evolution $S'_\mathbf{V}$ or $S_\mathbf{V}$ at different times, which are intimately connected with the network structure $\mathbf{V}$,  the bare frequencies of the network oscillators and the frequencies of the normal modes.  The evolutions described by Eq. (\ref{netoscEM}) and Eq. (\ref{netosc})  being  equivalent, we can choose one of the two pictures  according to which is more convenient for evaluating the desired expectation values or which one between $S'_\mathbf{V}$ and $S_\mathbf{V}$  could be more efficiently implemented in the experimental setup.
\begin{figure}
  \includegraphics[width=\linewidth]{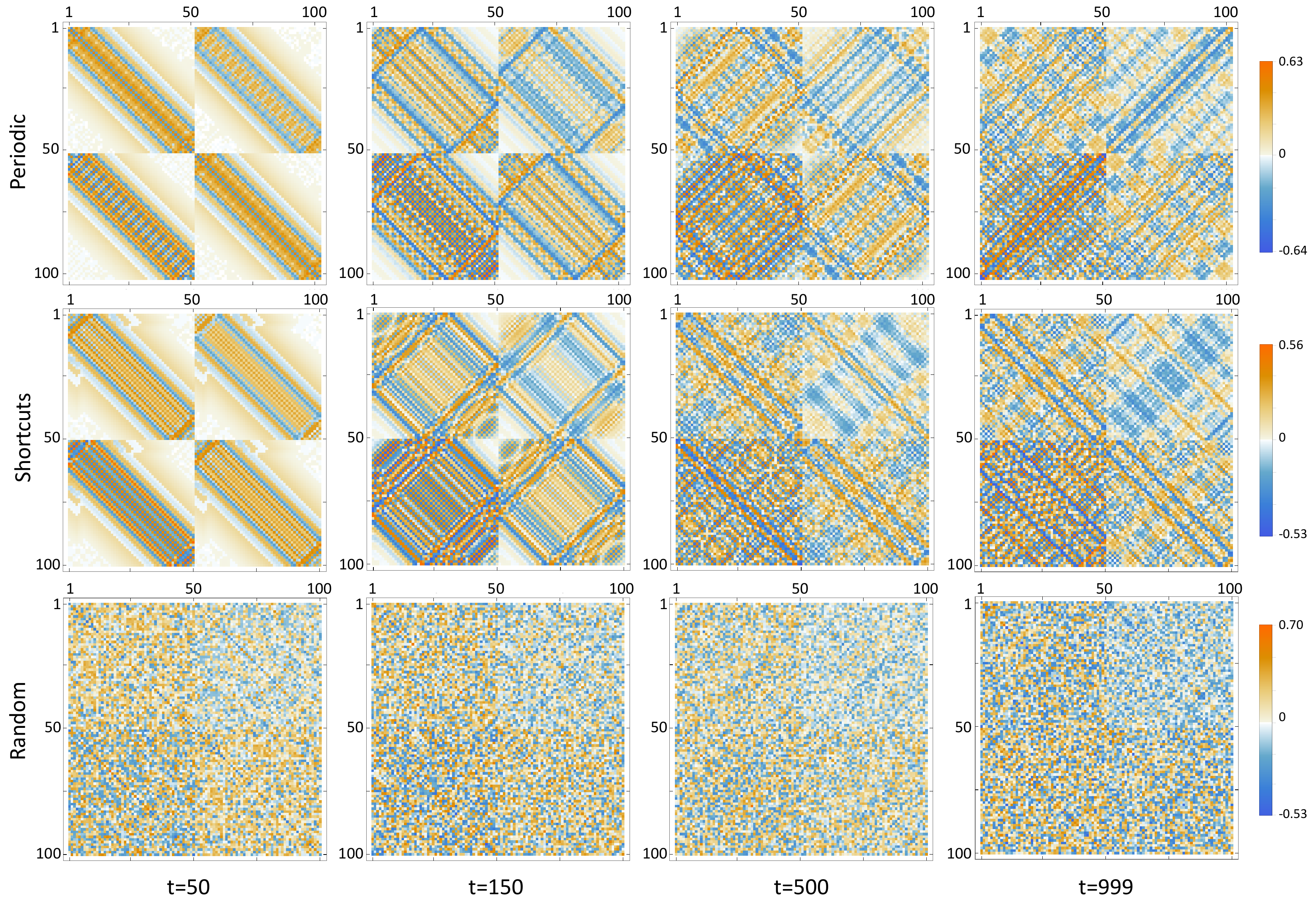}
  \caption{Matrix plot of the calculated  evolution  $S_\mathbf{V}$, i.e., the matrix of Eq. (\ref{netosc}), for the three networks of Fig.\ref{fig:net},  at different times, all the oscillators frequencies are set to $\omega_i=\omega_0=0.25$. The matrices contain $100$ (or $102$)  rows as they concern the evolution at the given time of the $50$ ($51$) $\mathbf{q}$ quadratures plus the $50$ ($51$) $\mathbf{p}$ quadratures of the network oscillators. A temperature colour map is used: positive values are in red colours, negative in blue while white colour indicates the zero value, as shown by the legend in the plot at $t=999$.
} 
  \label{fig:ev}
\end{figure}
The matrices $S'_\mathbf{V}$ and $S_\mathbf{V}$  in Eq. (\ref{Xx}) are symplectic and they can then be factored via the Bloch-Messiah (B-M) decomposition \cite{Braunstein05}, for example:
\begin{equation}
S_\mathbf{V}=R_1^T \Delta^{sq} R_2
\label{eq-bm}
\end{equation} 
  $\Delta^{sq}$ being a diagonal matrix  and $R_1$, $R_2$ symplectic and orthogonal matrices. 
 In Fig. \ref{fig:net} three different networks composed by  N=51 (Fig.\ref{fig:net} A) or N=50 (Fig.\ref{fig:net} B, C) nodes are shown: a periodic chain, a linear chain with shortcuts and a random network, the bare frequencies of the network oscillators are set to $ \omega_i=\omega_0=0.25$ for the three different topologies. The network matrices $S_\mathbf{V}$ calculated at different times from  the evolution in Eq. \ref{netosc} are shown in Fig. \ref{fig:ev}, while the B-M decomposition at time $t=50$ for the three network is shown in Fig. \ref{fig:BM}. \\
Note that the complexity of B-M decomposition is of the order of $O(N^3)$  \cite{DeGosson06}. 
 It takes around  $0.03$ second to implement the B-M decomposition for networks of 50 nodes with a standard laptop. 
 
 \section{Mapping the network to a multimode  experimental platform}

\begin{table*}[]
\centering
\label{Table}
\begin{tabular}{p{3.5cm}p{5.5cm}p{5.5cm}}
\hline
&Quantum network  & Experimental implementation \\
\hline
Node&Quantum harmonic oscillator & Optical mode\\
Link&Spring-like coupling term&Non-linear mode coupling\\
$t$&Evolution  time & Parameter controlling the symplectic matrix $S_{ex}$\\
Frequency&Oscillator frequencies&Squeezing of optical modes\\
Decoupled mode&Normal mode of the network & Supermode, a non-correlated squeezed optical mode\\
Addressing a node&Local measurement&Pulse shaping \& mode-selective measurement of the supemodes\\
\hline
\end{tabular}
\caption{The mapping of the quantum networks to the experimental platform. Refer to main text for more details.}
\end{table*}

The matrices $\Delta^{sq}$, $R_1$ and $R_2$  in eq. (\ref{eq-bm})  have a well-known physical interpretation in optics: the first is a squeezing operation while each of the two others corresponds to the action of a linear multiport interferometer, i.e. a basis change. The implementation of $S_\mathbf{V}$ can then be decomposed in a combination of squeezing and linear optical operations in a multi-mode scenario. In the following we are going to present a particular compact and completely reconfigurable experimental implementation of such strategy. 

The experimental implementation proposed here is based on parametric processes pumped by optical frequency combs. A very large network of entangled photons is produced \cite{Medeiros14,Roslund14} under a quadratic Hamiltonian, that leads to a symplectic evolution. The mapping we present here aims at matching this evolution to $S_\mathbf{V}$ using the experimentally controllable degree of freedoms: the pump spectral amplitude and the measurement basis. 

In practice, the laser sources considered here are mode-locked lasers emitting pulses from 30 $\mathrm{fs} $ to 150 $\mathrm{fs}$ with repetition rates from 70 $\mathrm{MHz}$ to 150 $\mathrm{MHz}$  at a wavelength centred around $\lambda=800\mathrm{nm}$. 
The spectrum of these lasers is constituted of hundreds thousands of frequency components. A second harmonic generation process is used for generating  a frequency-doubled comb which serves as pump of the parametric process.  The pump frequencies are written as $\omega_{p}=\omega_{p0}+p\cdot\omega_{RR}$, where $\omega_{RR}$ is the repetition rate of the original laser source, $\omega_{p0}$ is the pump central frequency and $p$ an integer. 
The Hamiltonian describing the parametric process, which is parametric down conversion in a $\chi^{(2)}$ crystal, is  $ H=\imath g \sum_{m,n}\mathcal{L}_{m,n}a^{\dag}_ma^{\dag}_n + h.c.$; it couples pairs of frequencies  $\omega_m$ and  $\omega_n$   via the term $\mathcal{ L}_{m,n}$ which is the product of crystal's phase matching function $f_{m,n}$ and of the complex pump  amplitude $\alpha_p$ at frequency $\omega_p=\omega_m+\omega_n$.  $\mathcal{ L}_{m,n}$ describes the probability amplitude that a photon at frequency $\omega_p$ is converted into two daughter photons at frequencies $\omega_n$ and $\omega_m$. Given the number of frequency components in the pump,  a number  of down-converted frequencies, which is of the same order of magnitude that the one in the original spectrum, is combined in a non-trivial multipartite entangled structure \cite{Gerke15}.  
\begin{figure*}
  \includegraphics[width=\linewidth]{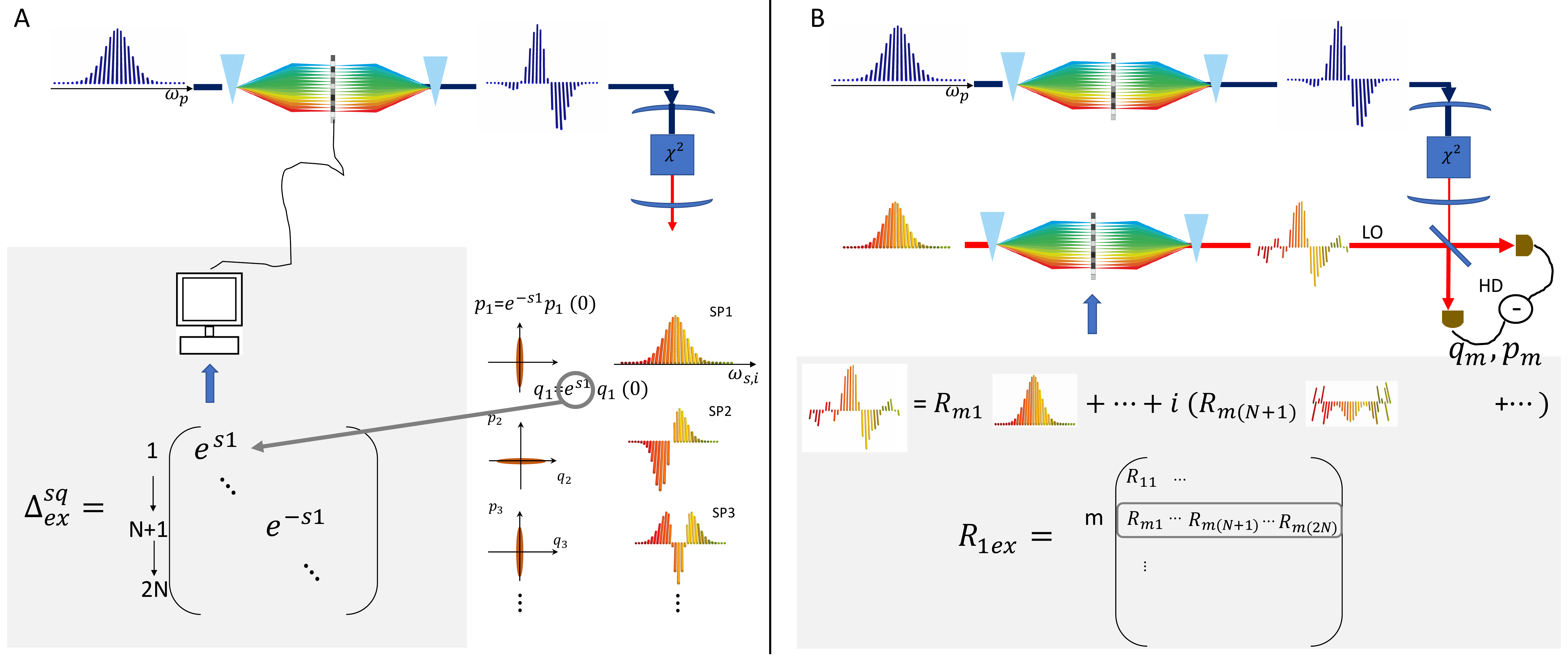}
  \caption{Experimental control  of the multimode squeezing operation $\Delta^{sq}_{ex}$ (A)  and the basis change $R_{ex}$ (B) . The first is controlled by  shaping the pump of the parametric process (here in cavity as \cite{Roslund14}), the experimental parameters of the pulse shaper are found by running an optimization procedure based on an evolutionary algorithm. The basis change defines the modes which carry the information on the oscillators evolution as a combination of the supermodes shape. }
  \label{fig:setup}
\end{figure*}
This Hamiltonian governs the system evolution that we express for the optical quadrature operators $\{ q_1, q_2,.. q_N, p_1,p_2,...p_N\}=\{\mathbf{q},\mathbf{p}\}=\mathbf{X}_b^T$ in a given mode basis $b$, with the convention
$q_j =(1/\sqrt{2})(a_j +  a^\dagger_j)$ and $\quad p_j = (i/\sqrt{2})( a^\dagger_j -  a_j)$. A natural choice for the basis would be the frequency components of the comb spectrum.
The parametric Hamiltonian leads to a symplectic transformation $\mathbf{X}_f=S_{ex}\mathbf{X}_i$ which can be decomposed via B-M in 
\begin{equation}
S_{ex}=R_{1ex}^T\Delta^{sq}_{ex}R_{2ex}
\end{equation}
If the input modes $X_i$ are in  the vacuum state the $R_{2ex}$ term can be neglected  
\begin{equation}\label{BMex}
\textbf{X}_f=R_{1ex}\Delta^{sq}_{ex}\textbf{X}_i.
\end{equation}
 The squeezing parameters in $\Delta^{sq}_{ex}$ can be derived from the eigenvalues of $\mathcal{ L}$ and $R_{1ex}$ can be reinterpreted as change of basis from the basis of the squeezed modes, named 'supermodes', towards  the measurement basis. 
 In the case of a pump  and phase matching function with Gaussian spectral profile, the supermodes basis obtained from $\mathcal{L}$  coincide with Hermite-Gauss modes.  
  If the symplectic transformation $S_{ex}$ can be experimentally controlled in order to coincide with  $S_\textit{V}$  we are obtaining an optical implementation of the complex network  described by the equation (\ref{netosc}).  This means that by addressing the optical modes we can measure the properties of the oscillators in the network.
  
We would like to stress here the special features of our mapping: to emulate the network dynamics we only need to know its symplectic evolution  $S_\textit{V}$ at time t, which can be experimentally emulated by  $S_{ex}$. We don't need to know the physical phenomenon that generates the networks dynamics or the nature of the oscillators in the network (they could be mechanical or optical oscillators) or even the specific picture that can be chosen to describe the Hamiltonian evolution. In the case described by eq. (\ref{netosc}), for example, the symplectic evolution is derived from the full Hamiltonian also containing the non-interacting terms, while the experimental process is described in the interaction picture. As a consequence, the frequencies of the network oscillators will not match the frequencies of the correspondent modes, but instead will be mapped to squeezing parameters in the optical setup.  By focusing on the emulation of the symplectic evolution, we can then use the same experimental resources to map different network parameters according to the features of the phenomenon we are studying. \\
The mapping of the quantum networks to the experimental platform is summarized in Table 1.
\begin{figure*}[ht]
  \includegraphics[width=\linewidth]{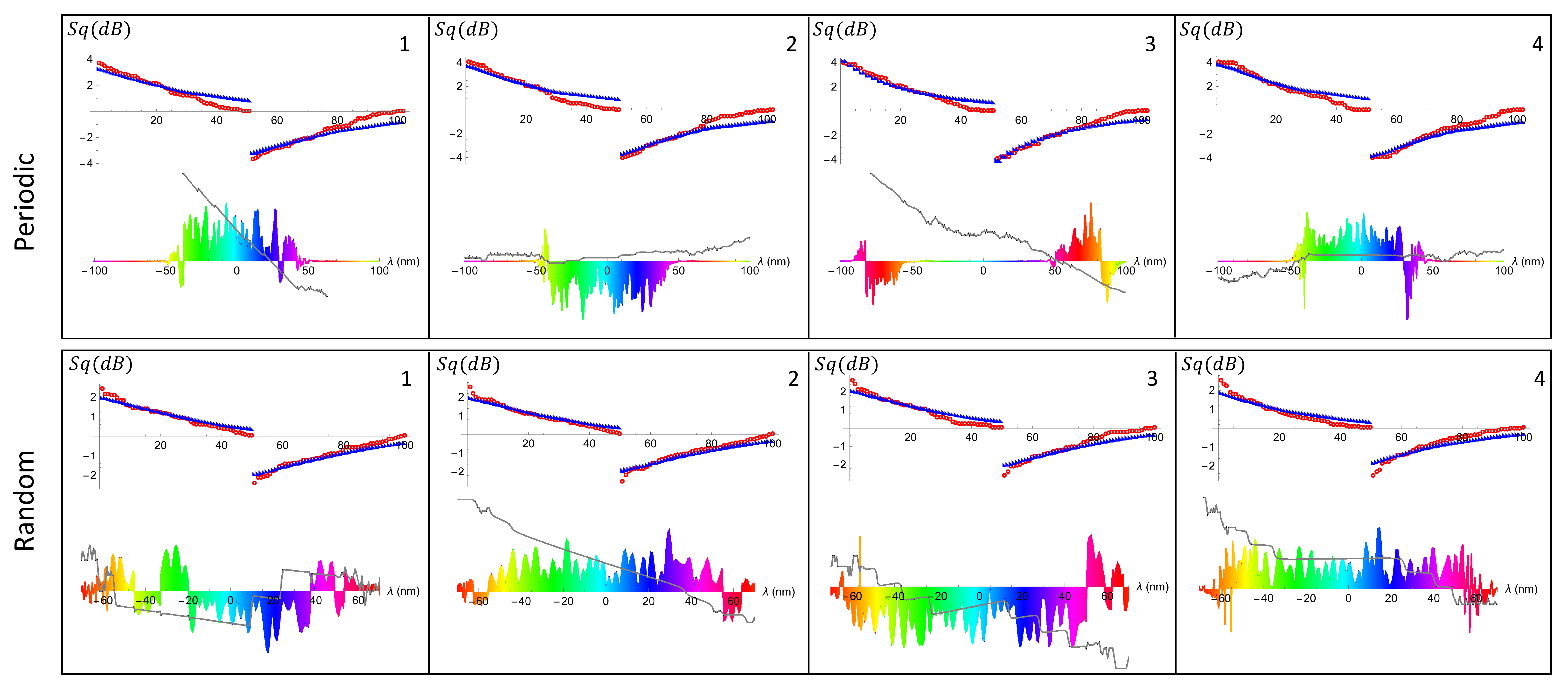}
  \caption{Simulation of the periodic and the random network evolution by the experimental setup.  1,2,3 and 4 stand for t=50, 150, 500 and 999 respectively. Upper part of each box: diagonal terms of the matrix $\Delta^{sq}$ in dB, the red are the theoretical values coming from the B-M decomposition, the blue are the optimized values which can be obtained from an optimized pump shape with a BiBO crystal of length $2.5$ mm, in the case of the periodic network, and of $1.5$ mm, in the case of random network.   Lower part of each box: spectral shape of mode $26$ representing the $26^{th}$ oscillator obtained by applying the linear transformation $R_1^T$ on the calculated supermodes. The amplitude is represented by a plot with rainbow colors while the gray line is the spectral phase. Units are arbitrary. This corresponds to the spectral shape which has to be set in the Local Oscillator (LO) in order to address the mode.}
  \label{fig:sim-ra-pe}
\end{figure*} 

\begin{figure}[ht]
\begin{center}
  \includegraphics[width=8cm]{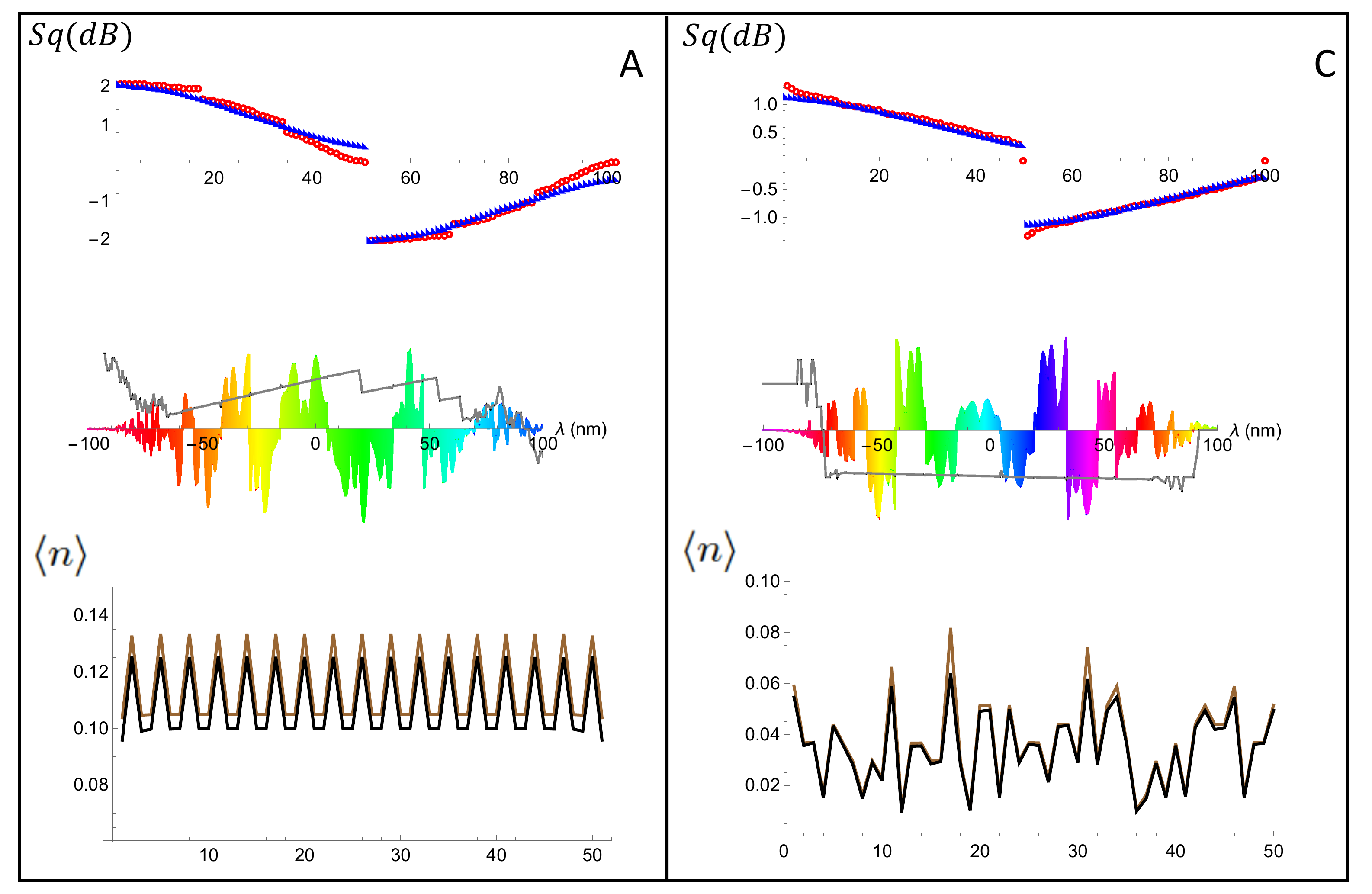}
  \caption{Simulation of the periodic (A) and the random (C) network in the eigenmode picture at t=50 with a crystal length of $1.5$ mm.  The squeezing values here are the diagonal terms of $\Delta^{sq}$ from the B-M decomposition of $S'_\mathbf{V}$  in Eq. (\ref{Xx}). In the center: the pulse shape for the $26^{th}$ node. In the lower part the calculated mean photon number of each oscillator (brown curve) and the simulated one (black curve) from the experimentally achievable squeezing values.}
  \end{center}
  \label{fig:sim-n}
\end{figure}

 In the following we explain how the symplectic transformation can be controlled by manipulating the terms of its B-M decomposition. 
The squeezing operations in $\Delta^{sq}_{ex}$ can be controlled by tailoring the spectral shape of the pump in the parametric process. This can be done experimentally via a pulse shaper based on a Spatial Light Modulator (SLM) in a 4-f configuration (see figure \ref{fig:setup})  which provides control over amplitude and phase of the spectral components by exploiting a 2-D geometry . The phase mask of the SLM can be set according to the results provided by an optimization procedure, based on an evolutionary algorithm which directly relates experimental parameters with the diagonal terms of the matrix  $\Delta^{sq}_{ex}$\cite{Arzani17}. Note that the optimization of $\Delta^{sq}_{ex}$ also influences the spectral shape of the supermodes, and thus the squeezed modes basis. 

The linear transformation  $R_{1ex}$ can be tuned adequately choosing the measurement basis, which is the equivalent of choosing the basis on which evolution (\ref{BMex}) is calculated. Experimentally, this is implemented using homodyne detection where the measured mode is governed by the spectral shape of the Local Oscillator (LO)\cite{Cai17}.  The squeezing matrix $\Delta^{sq}$ acts on the supermode basis, then the quadratures of the \textit{m-th } network oscillator can be addressed  by applying the basis change defined by the \textit{m-th } row of the matrix $R^T_{1}$. This means that we can measure the quadratures  $q_m(t)$ and $p_m(t)$ of the  \textit{m-th } network oscillator by performing homodyne detection with a LO shape defined by $u_{LO-m}(\omega)=\sum_{j=1}^N R^T_{1mj}SP_j(\omega) +\imath \sum_{j=N+1}^{2N} R^T_{1mj}SP_{j-N}(\omega) $,  where $SP_i(\omega) $ are the supermodes. 
 Figures \ref{fig:setup}  A and B show the scheme for the experimental control of  $\Delta^{sq}_{ex}$  via the control of the pump spectral shape, after the optimization, and the control of  $R_{1ex}$ by choosing the measurement basis via pulse shaping of the LO. 
 
A second point has to be stressed here: even if the B-M reduction of general symplectic transformations and its physical interpretation is well known, the full control allowing for complete reconfigurability of the squeezing operations and basis changes in a single optical setup, enabling from trivial to complex topologies, has never been addressed before.
 \begin{figure*}[ht]
\begin{center}
  \includegraphics[width=12cm]{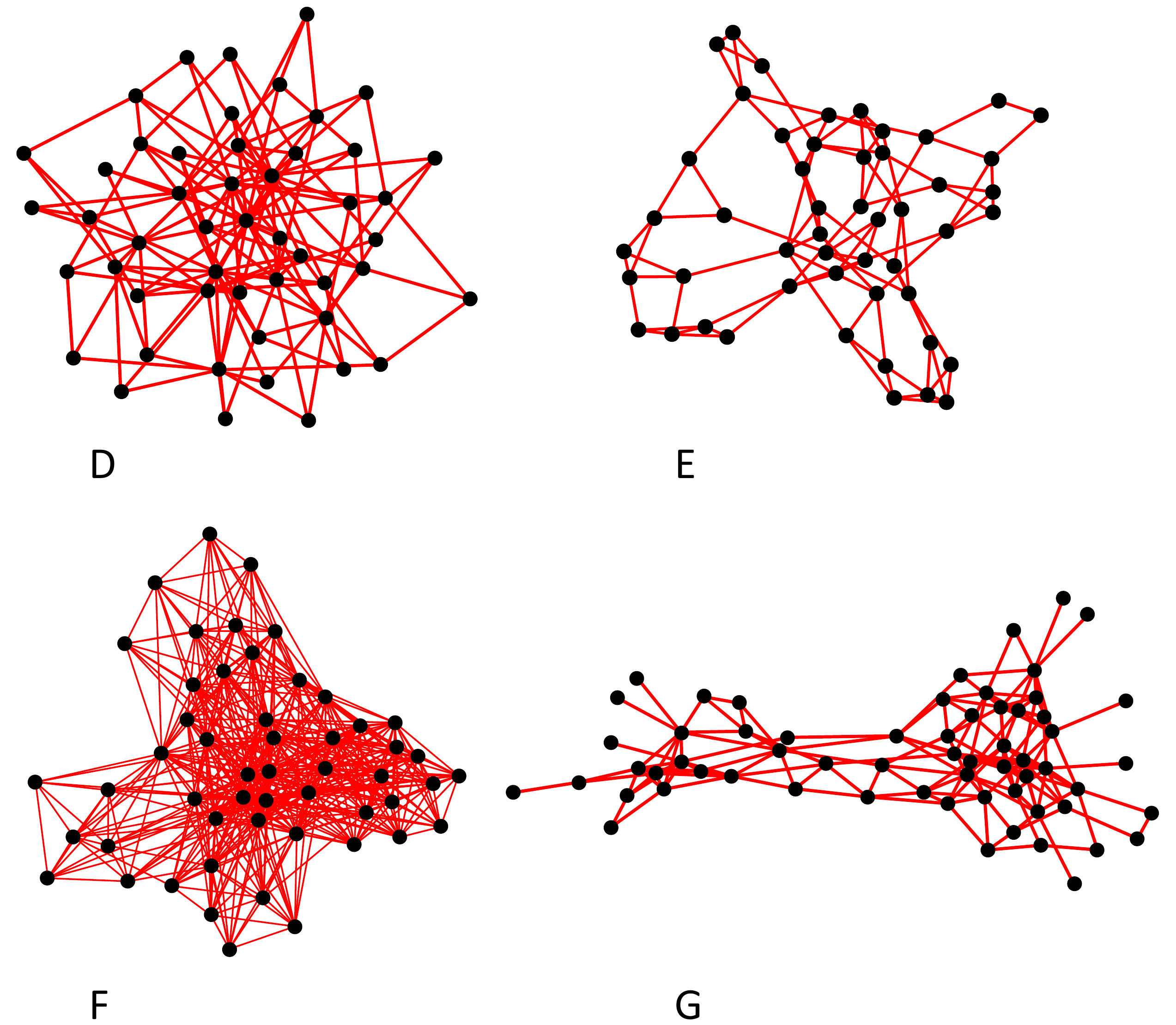}
  \caption{Graphical structure of supplementary networks. \textbf{D} is a 50-nodes network derived from the Barab\'{a}si-Albert model \cite{Barabasi99} , characterized by a power law distribution of the degree with connection parameter set to $3$. \textbf{E} is a 50-nodes net  work derived from the Watts–Strogatz model \cite{Watts98}, characterized by short average path lengths and high clustering, the rewiring probability is set to be $0.2$.  \textbf{F} and \textbf{G} are networks retrieved from a public repository of real-world complex networks. \textbf{F} is a 52 nodes-network which represents the  connectional organization of the cortico-thalamic system of the cat, as studied by J.W. Scannell et \textit{al.} \cite{Scannell99}. \textbf{F} is a 62-nodes network representing the social network of dolphins, derived by D. Lusseau \cite{Lusseau03}. The coupling strength between the nodes for the four networks is set to be  $v_{ij}=0.05$. }
  \label{fig:net2}
 \end{center} 
\end{figure*}

  \begin{figure*}[ht]
  \includegraphics[width=\linewidth]{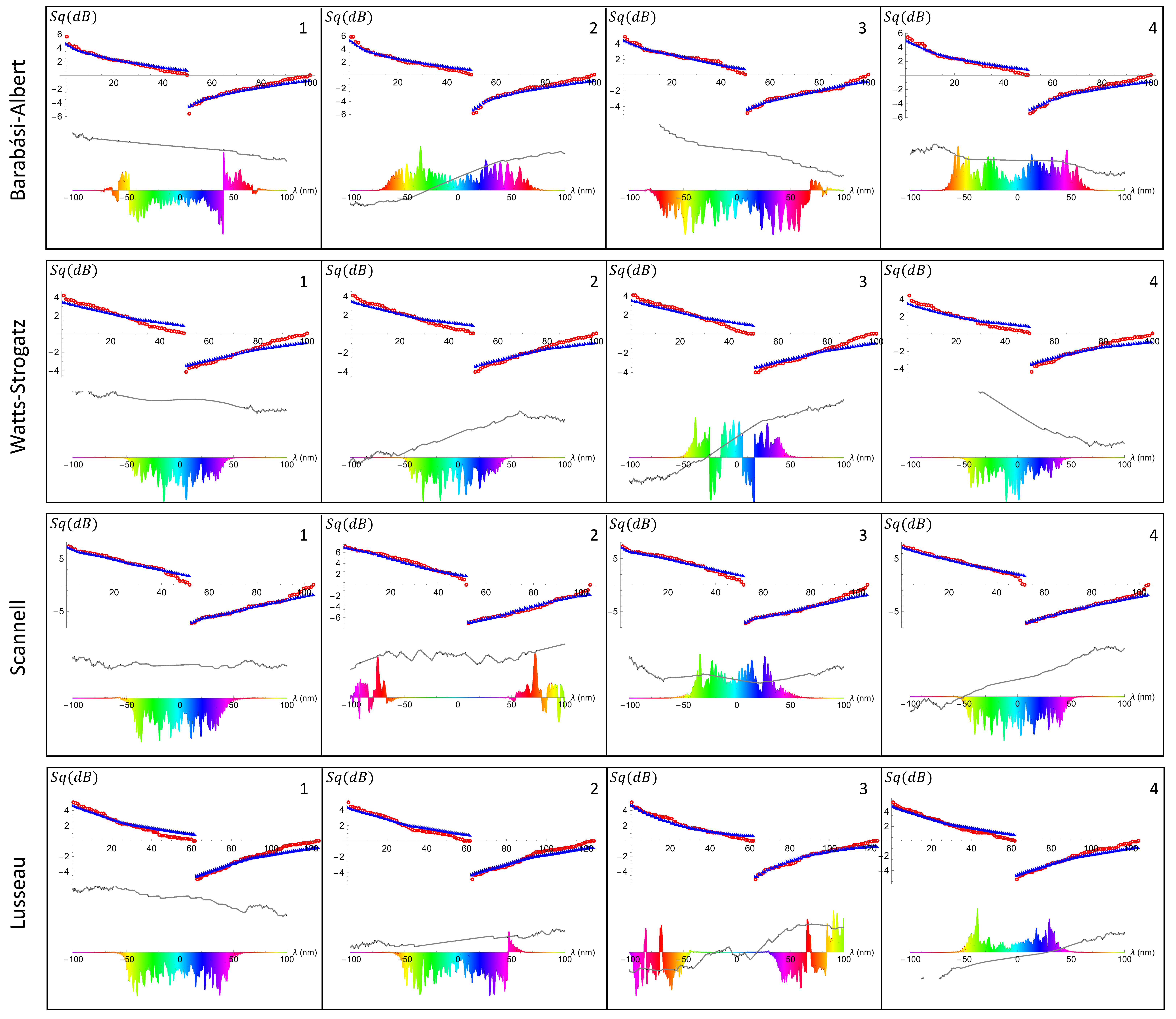}
  \caption{Simulation of the supplementary networks shown in Fig. \ref{fig:net2}. The optimization is performed  with a BiBO crystal of length $1.5$ mm  in the case of the Barab\'{a}si-Albert network, and of $2.5$ mm for the other three networks. }
  \label{fig:sim-others}
\end{figure*} 
The control of $R_{ex}$ and  $\Delta^{sq}_{ex}$ is sufficient for the simulation of the dynamics if  the particular feature we are interested in is independent on the initial quantum state of the network oscillators.  In that case we can consider the oscillators being in the vacuum state at time $t=0$, by choosing  $R_{ex}=R_1^{T}$  $\Delta^{sq}_{ex}=\Delta^{sq}$, and discarding $R_2$  the network simulation can be performed.  Otherwise  we need to initialize the optical modes in the desired quantum states. The oscillators can be prepared in an initial pure Gaussian state, by considering a first symplectic transformation $S_0$ which prepares the modes in the desired state, followed by network evolution $S_\mathbf{V}$. The effective total evolution to be simulated is 
\begin{equation}
\mathbf{x}(t)=S_\mathbf{V} S_0\mathbf{x}(0)=S_{ef}\mathbf{x}(0)
\end{equation}
 where the initial states are vacua. Then via B-M decomposition we can set
 \begin{equation}
 S_{ef}=R_{1ef}\Delta^{sq}_{ef}R_{2ef},
 \end{equation}
 experimentally implement $\Delta^{sq}_{ex}=\Delta^{sq}_{ef}$  and $R_{ex}=R_{1ef}$ and finally discard $R_{2ef}$ as it is acting on vacuum states. 
 If the complex networks we are implementing are intended to simulate environments of open quantum systems, a typical requirement we may have to fulfill is to set a finite temperature for the network.  This can be achieved by initializing the oscillators in an optical thermal state, which can be built, for instance, as  the reduced state of an entangled state. Indeed, a maximally entangled state of 2 qubits has marginal distributions which are maximally mixed (equivalent to $T\to\infty$), and the same is true in
continuous variable Gaussian states. Consider e.g., a two-mode squeezed state of two bosonic modes with squeezing parameter $r$. The reduced covariance matrix
of only one of the modes  reads $\sigma=\textrm{diag}\{ \hbar\omega \cosh(2r)/2,\hbar \cosh(2r)/2\omega \}$, which describes a thermal state with 
$\cosh(2r)\rightarrow\coth(\hbar\omega/2T)$. For $T/\omega>1$ the required squeezing grows as $r\simeq \textrm{acosh}(2 T)/2$ and this scales like $\log T$, e.g. $r\sim 2$ for $T/\omega\sim 20$, 
which is reasonable experimentally.
We may want to initialize one oscillator of the network in a thermal state. This is then done by having two probes initially in a two-mode squeezed state, which is obtained by a $\pi/4$ rotation in the $2\times 2$ subspace, individual squeezings, and a rotation back.
If instead we need to initialize the entire network into a thermal state we should duplicate it and impose a two-mode squeezed state between each pair of node copies. Network dynamics then follows by applying the time evolution only on half of the network units, and the identity on the other half. This scheme increases the overhead of the protocol but is feasible in principle.

The Figure  \ref{fig:sim-ra-pe}  shows  the results of the simulation for the periodic network and the random network based on experimental parameters. The optimization procedure for the pump shape is run in order to minimize the distance between the theoretical values (the diagonal of $\Delta^{sq}$) and the experimentally realisable ones ($\Delta^{sq}_{ex}$) by varying the spectral shape of the pump. 
The simulated squeezing values are close to the theoretical ones: the relative distance of the simulated curves to the theoretical ones are in between $6.8\%$ and $10.4\%$ for the periodic network and in  between  $3.5\%$ and $5.3\%$ for the random network.  The spectral shape of the mode corresponding to one oscillator of the network is given: 0 nm corresponds to the central wavelength of the laser
source (800 nm).  The relative errors on the experimentally simulated $S_\mathbf{V}$ are exactly the same of  $\Delta^{sq}$ if the exact required basis change $R_1$ is performed. This necessitates that any of the 50 modes simulating the oscillators can be addressed by choosing the right shape of the LO in homodyne detection. As shown in Fig. \ref{fig:sim-ra-pe}, this requires to handle laser light of large bandwidth, which can be generated by coherent broadening of the main laser source.
The results of Fig. \ref{fig:sim-ra-pe}  are obtained by setting two  different crystal lengths: in the case of the shorter crystal ($1.5$ mm) the experimentally achievable squeezing matrix ($\Delta^{sq}_{ex}$) is closer to the theoretical one ($\Delta^{sq}$) than in the case of $2.5$ mm crystal length. However the case of shorter crystal  generally requires larger bandwidth for the LO as the resulting supermodes have a larger spectrum. 
In Fig. \ref{fig:sim-n} we show one observable, the mean photon number, for all the oscillators in the network. We evaluate both its theoretical value with the exact evolution and the simulated one with the squeezing values derived from the optimization. For this calculation we choose the vacuum state as the initial state. The relative errors on the calculated excitation numbers is $5.3\%$  for the periodic and $5.0\%$  for the random network. The simulation is performed with a crystal length of $1.5$ mm, also in this case by admitting larger relative errors (of the order of $10\%$ in the excitation numbers) it is possible to reduce the bandwidth of the LO (to half the value of the shapes shown in Fig. \ref{fig:sim-n}) with a crystal length of $2.5$ mm. 

In order to show that the setup has the potential of replicating the behaviour of  complex networks with any topology by simulating $S_\mathbf{V}$ or $S'_\mathbf{V}$ at different times, we repeat the procedure for other networks with more complex topology, which are shown in Fig. \ref{fig:net2}.  We study the Barab\'{a}si-Albert model \cite{Barabasi99}, characterized by a power law of the degree distribution, and the Watts–Strogatz model \cite{Watts98}, which describes networks with small-world properties as clustering. Additionally, we consider two graphical structures derived from real natural networks,  studied by J.W. Scannell et \textit{al.} \cite{Scannell99} and D. Lusseau \cite{Lusseau03}, that we call Scannel and Lusseau networks. The simulations are shown in Fig. \ref{fig:sim-others} and they show the same performances of the simulations in Fig. \ref{fig:sim-ra-pe} with a similar trade-off between  relative errors limitation and reduction of  the LO bandwidth.

\section{ Interaction with additional oscillators}
In a more general scenario we would like to describe the network interacting with additional oscillators: these would, for example, depict the principal quantum systems which interact with the bath described by the network or the systems in which the information we want to convey through the network is initially encoded. We take $M$ oscillators with frequencies $\{\omega_{S1},\omega_{S2},.. \omega_{SM}\}$,  each of them interacting with one node according to the Hamiltonian $H_{Ir}=-kq_{Sr} q_i/\sqrt{\omega_{Sr}\omega_i}$.  The interaction couples the supplementary oscillators with any of the normal mode, as we can see from the interaction Hamiltonian in the decoupled quadrature picture: $H_{Ir}=-kq_{Sr}\sum_j\mathbf{K}_{ij}Q_j/\sqrt{\omega_{Sr}\Omega_j}$. The new total Hamiltonian again can be written in  diagonal form by orthogonal diagonalization of  $\mathbf{B}=\mathbf{O}(\mathbf{\Delta}_f/\sqrt{2})^2\mathbf{O}^T$ , the matrix analogue to A  containing  the extra terms given by the interaction.  $\mathbf{\Delta}_f$ is a diagonal matrix $\textup{diag}\{f_1,...,f_{N+M}\}$. 
\begin{equation}
H_{S+E+I}=\sum_j^{N+M}f_j(\mathcal{P}_j^2+\mathcal{Q}_j^2)/2.
\end{equation}
It is convenient to  describe the dynamics of the  network plus  the  new oscillators in terms of the evolution of the quadratures for the network eigenmodes and the supplementary systems 
$\mathbf{X}^T=\{\mathbf{Q}, \mathbf{q}_S, \mathbf{P}, \mathbf{p}_S\}=\{\mathbf{Q}, q_{S1},..,q_{SM}, \mathbf{P}, p_{S1},..,p_{SM}\}$ 
\begin{equation}\label{evolution}
\begin{pmatrix}
\mathbf{Q}(t) \\
\mathbf{q}_S(t) \\
\mathbf{P}(t) \\
\mathbf{p}_S(t)
\end{pmatrix}  = 
\begin{pmatrix}
O_1 D_{\cos} O_1^{-1} & O_1 D_{\sin} O_2^{-1} \\
-O_2 D_{\sin} O_1^{-1} & O_2 D_{\cos} O_2^{-1}
\end{pmatrix}
\begin{pmatrix}
\mathbf{Q}(0) \\
\mathbf{q}_S(0) \\
\mathbf{P}(0) \\
\mathbf{p}_S(0)
\end{pmatrix} 
\end{equation}
where we  introduced the diagonal matrices $D_{\cos ii}=\cos(f_i t)$ and $D_{\sin ii}=\sin(f_i t)$, and the  matrices $O_1=\sqrt{\mathbf{\Delta}_{\Omega,\omega_S}}\mathbf{O}\sqrt{\mathbf{\Delta}_{f}^{-1}}$ and $O_2=\sqrt{\mathbf{\Delta}_{\Omega,\omega_S}^{-1}}\mathbf{O}\sqrt{\mathbf{\Delta}_{f}}$, being
$\mathbf{\Delta}_{\Omega,\omega_S}=\textup{diag}\{\Omega_1,...,\Omega_N,\omega_{S1}...\omega_{SM}\}$. 
Equation  (\ref{evolution}) can be again written in a more compact form as  $\mathbf{X}(t)=S\mathbf{X}(0)$ with $S$ a function of time. 
The simulation of the dynamics of the network interacting with the system can then be performed by experimentally addressing a collection of oscillators $\mathbf{X}$ and implementing the transformation $S$  at any time.

\section{Probing the spectral density of a structured environment}
The additional oscillators introduced above can also be considered as probes for the network in the framework of the theory of open quantum systems \cite{Davies76,Breuer07}. Here, the probes would be considered the open quantum systems and the network their environment, and the objective would be to deduce some properties of the environment based on knowing the reduced dynamics of the probes. In the following we consider the case of a single probe and we describe the simulation, based on experimentally realisable operations, of the probing  technique for the network spectral density as proposed in \cite{Nokkala16}.

The spectral density $J(\omega)$, a key quantity in open quantum systems theory, is determined by the environment and interaction Hamiltonians $H_E$ and $H_I$, and gives the density of environmental modes as well as coupling strengths to them as a function of frequency \cite{Breuer07,Vasile14}. Nothing else about $H_E$ and $H_I$ needs to be known to determine the open system dynamics. Conversely, knowing the open system dynamics reveals information about $J(\omega)$. Indeed, provided that the coupling between the probe and the environment is sufficiently weak, measuring the photon number operator  $\langle n\rangle $ of the probe having frequency $\omega_{S}$ allows to approximate $J(\omega_{S})$ from

\begin{equation}\label{J}
J(\omega_S)=\frac{\omega_{S}}{t}\ln\left( \dfrac{N(\omega_S)-\langle n(0)\rangle  }{ N(\omega_S)-\langle n(t)\rangle} \right),
\end{equation}
where $N(\omega_S)=(e^{(\omega_S/T)}-1)^{-1}$ is the thermal average boson number and $t $ is a large enough interaction time.In the case at hand, one may couple the probe weakly to any subset of network oscillators and use Eq. (\ref{J}) to probe $J(\omega_S)$, however different subsets correspond to different spectral densities as the function depends also on $H_I$.

The mean photon number here can be accessed from the homodyne measurement of the probe as $\langle n(t)\rangle=1/2\langle q_S(t)^2\rangle+1/2\langle p_S(t)^2\rangle -1/2$. The quadratures of the probe $\{q_S(t),p_S(t)\}$ appear in the left hand side of the Eq. (\ref{evolution}), which can be used to derive their expectation value at any time.
We have to stress here that when the initial mode basis $\mathbf{X}(0)$ is defined, the B-M decomposition of $S$ (the matrix appearing on the left side of Eq. (\ref{evolution})) completely identifies, via the basis changes,  the final mode corresponding to 
$\{q_S(t),p_S(t)\}$, which can be simply addressed by pulse-shaped homodyne detection. In this case we cannot initialize all the network nodes and the probe in the vacuum state, because no energy exchange between the network and the system would be mapped  in $\langle n\rangle$.  We then will use a first operation able to set a non-zero value for  the mean photon number of the probe $\langle n(0)\rangle$, as for example a squeezing operation $S_0=\Delta'$, and then we will implement the B-M decomposition of $S_{eff}=S\Delta'$  \\
The protocol is the following one: given a network structure, a probe with frequency $\omega_S$ is chosen and the evolution of the total system is computed at an appropiate time \textit{t}, the experimental parameters are set according to the calculated B-M giving $R_{1ef}, \Delta^{sq}_{ef}$. The probe mode is measured by pulse shaping the LO and one value of $J(\omega_S)$ is obtained. The procedure is repeated for other $\omega_S$ in order to retreive the full shape of the spectral density.

\begin{figure}
\begin{center}
  \includegraphics[width=\linewidth]{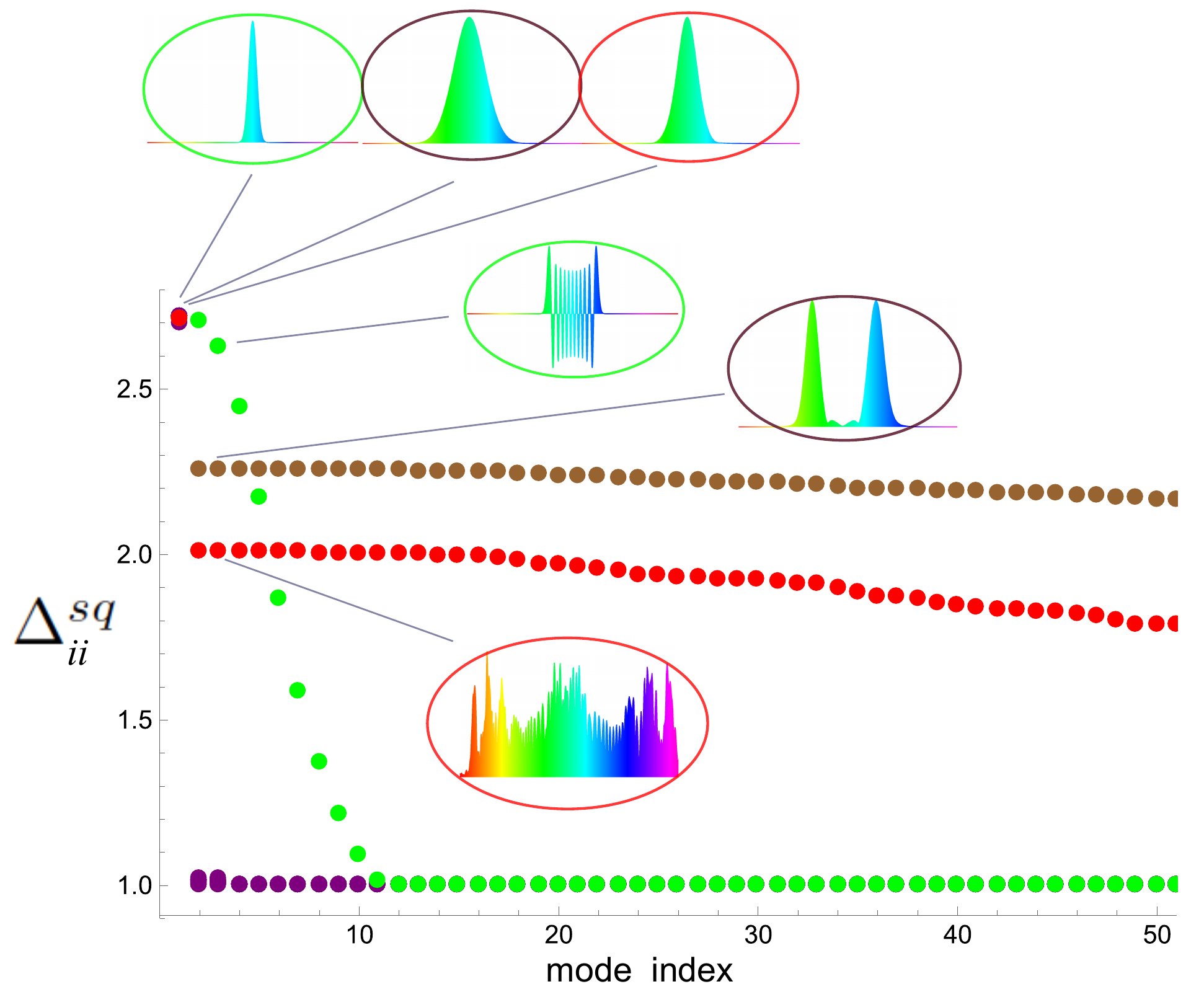}
  \caption{Purple: diagonal elements of the target $\Delta^{sq}_{ef}$; green: experimental achievable squeezing spectrum, i.e. diagonal elements of Kex in the case of a Gaussian pump spectrum for the parametric process with a crystal length of $1.5$ mm and with no  pump optimization. 
  Brown and red:  experimental achievable squeezing spectrum, when optimization on the pump shape is performed with a crystal length of $0.5$ mm and $1.5$ mm.  In the three cases (green, red and brown) the first diagonal element is exactly the same as the target one (purple). The pictures in the circles show the spectral shape of the first and the third supermode in the three possible experimental configuration.}
  \label{fig:sq}
\end{center}
\end{figure}
\begin{figure}
  \includegraphics[width=\linewidth]{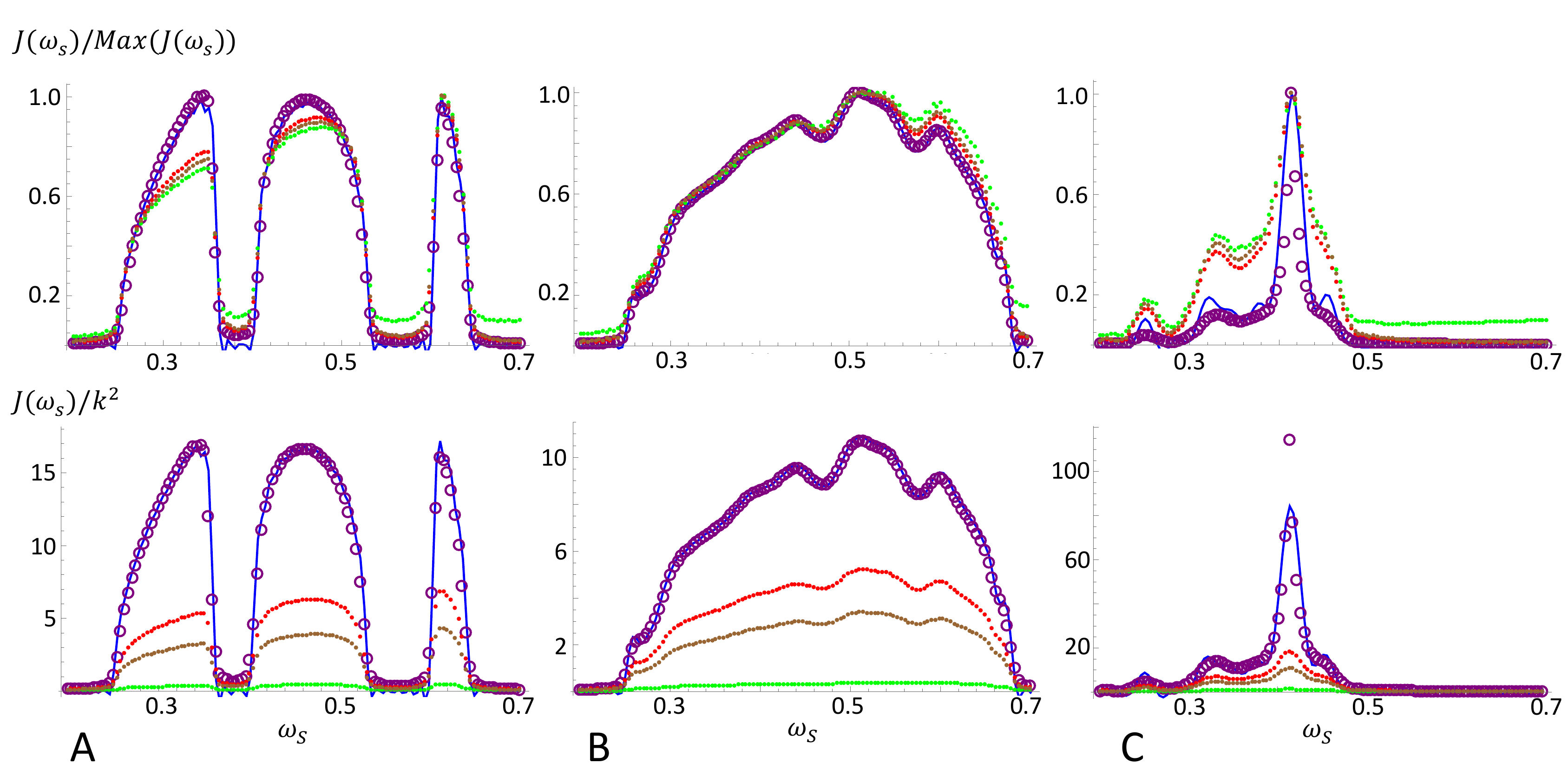}
  \caption{The spectral density  $J(\omega)$ for the periodic (A), the shortcuts (B) and the random network (C).  
Blue line: analytical calculation,  violet circles: exact probing method. Green dots: calculation of  $J(\omega)$ by replacing  $\Delta^{sq}_{ef}$ with the $\Delta^{sq}_{ex}$ which can be obtained in the experimental setup with no pump optimization. The red and the brown dots are calculated with the  $\Delta^{sq}_{ex}$  obtained from pump optimization with a crystal of $1.5$ mm and $0.5$ mm. Top: the calculated  spectral density is plotted after normalization to its maximal value. Down: the results of direct calculation are shown. }
  \label{fig:jw}
\end{figure}

\begin{figure}
\includegraphics[width=\linewidth]{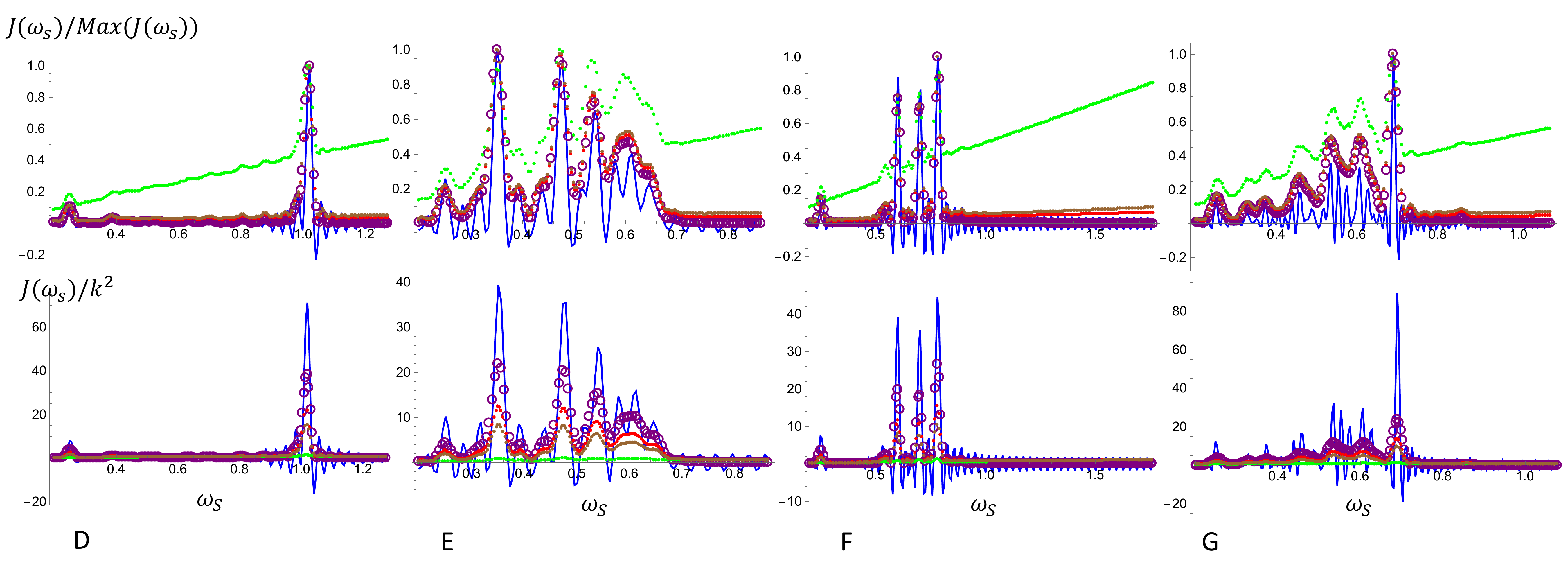}
  \caption{The spectral density  $J(\omega)$ for the Barab\'{a}si-Albert  (D), the Watts–Strogatz (E),  the Scannell (F)  and the Lisseau (G) networks. Shapes and colors have the same meaning of Fig. \ref{fig:jw}}
  \label{fig:jw2}
\end{figure}

We now look at how the network and probe parameters are mapped in the model and particularly in the B-M decomposition of the symplectic matrix in Eq. (\ref{evolution}). The network is totally identified by the matrix $\mathbf{A}$ , which contains the geometrical structure, the frequency of the  oscillators and the strength of the couplings between them. The probe is identified by its frequency $\omega_S$ and the strength of its coupling $k$ to one of the network node which is supposed to be weak in the probing scheme. 
For simplicity we set a constant value for the frequencies and the couplings in the network, i.e.  $\omega_i=\omega$ and $v_{ij}=g$.  If we fix  the set of values $\{\omega, g, \omega_S, k\}$ we find almost the same matrix $\Delta^{sq}$ in the B-M decomposition of $S$ for any network at any time, while the different network geometries lead to completely different basis changes  $R_1,R_2$.
The frequency terms $\{\omega,  \omega_S\}$ in fact enter in the renormalized quadrature definitions and they contribute in the Hamiltonian as squeezing terms, while the parameters  $\{ g,  k \}$ are the ones driving the quadratic Hamiltonians which contain squeezing operations. On the other hand the basis change transformations have almost the same role as in cluster state generation, by defining the graphical structure, and they are hence intimately connected with the network structure.
When the probe is in a very weak coupling regime $k<0.01$, condition which can be always chosen without affecting the results of the probing technique, the influence of $\omega_S$ on the squeezing matrix $\Delta^{sq}$ is very small. This  simplifies the experimental procedure of the probing protocol because  scanning $\omega_S$ requires almost no changes in the squeezing structure (i.e. in pump shaping) but only appropriate pulse shaping of the LO mode, according the updated basis change.
 Fig. \ref{fig:sq} shows in purple the target squeezing, i.e. the diagonal elements of $\Delta^{sq}_{ef}$,  in the case of a periodic chain, with parameters set as in Fig. \ref{fig:net};  ten different purple lists are superposed corresponding to ten different values of $\omega_S$.  The configuration in which we probe with an initially squeezed oscillator requires only one significantly squeezed mode in the matrix $\Delta^{sq}_{ef}$, this is the reason why all the 10 purple lists almost coincide. The diagonal elements of three  experimentally achievable  $\Delta^{sq}_{ex}$ are shown: the green points correspond to a Gaussian profile for the pump of the parametric process (no pump shaping), while the brown and red points are obtained from the optimization procedures run considering a crystal length of $0.5$ and $1.5$ mm. In the three cases the first mode can be set to have exactly the required value of squeezing of the first target mode.  The insets show the shape of the first and the third supermodes corresponding to the experimentally achievable $\Delta^{sq}_{ex}$.
 In Fig.  \ref{fig:jw}  and  \ref{fig:jw2} we compare the results for the analytically calculated  $J(\omega)$ (blue lines), the one derived from the exact probing (violet circles), and the ones calculated by replacing the exact $\Delta^{sq}_{ef}$ with the experimentally achievable $\Delta^{sq}_{ex}$ (green, red and brown dots).
 We show the normalized value of $J(\omega)$  (top) and the raw values after  applying Eq. (\ref{J}). The experimentally achievable configurations allow for the retrieval of the peculiar shape $J(\omega)$ for the seven kinds of network. 
 In the case of no optimization of the pump shape, even if the intrinsic structure of the spectral density is mainly conserved, the numerical values of the calculated $J(\omega)$ are quite low, while larger values are obtained  when optimization via  pump shaping is included. 

\begin{figure}
  \includegraphics[width=\linewidth]{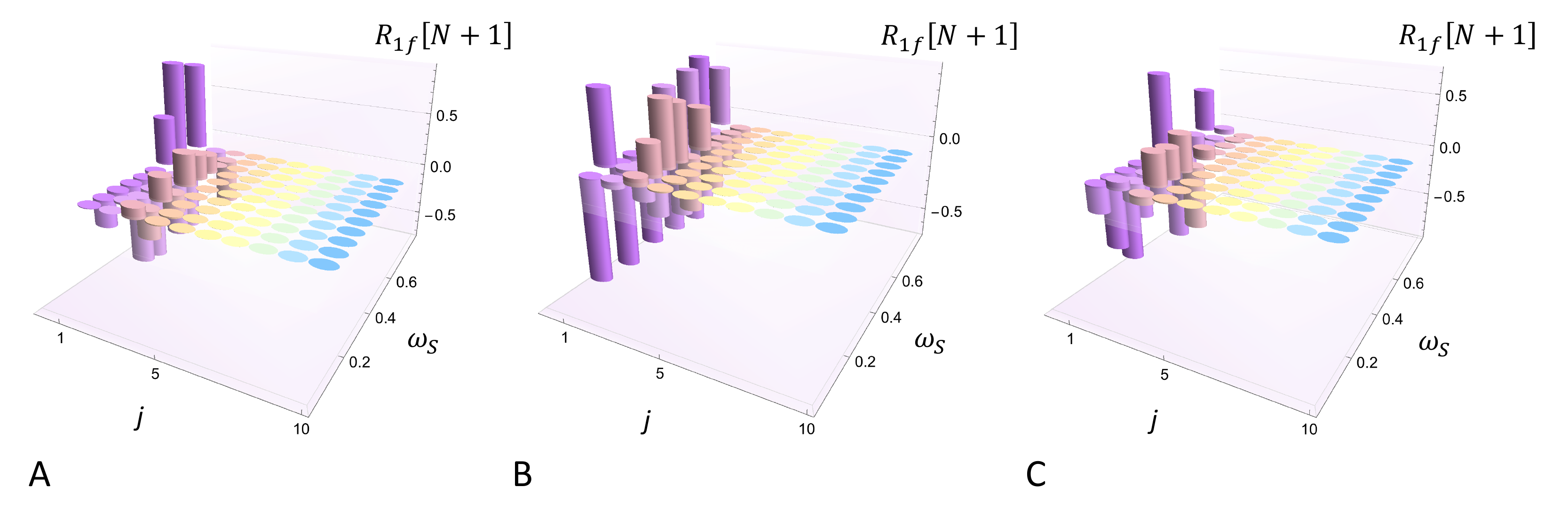}
  \caption{ Calculated first 10 elements of the N+1-th row of  $R_{1f}$, defining  the shape of the probe mode  as a linear combination of the initial supermodes, for A: periodic B: shortcuts and C: random network.
 They define the shape of the LO  as shown in Fig. \ref{fig:setup} B. }
  \label{fig:R}
\end{figure}

\begin{figure}
  \includegraphics[width=\linewidth]{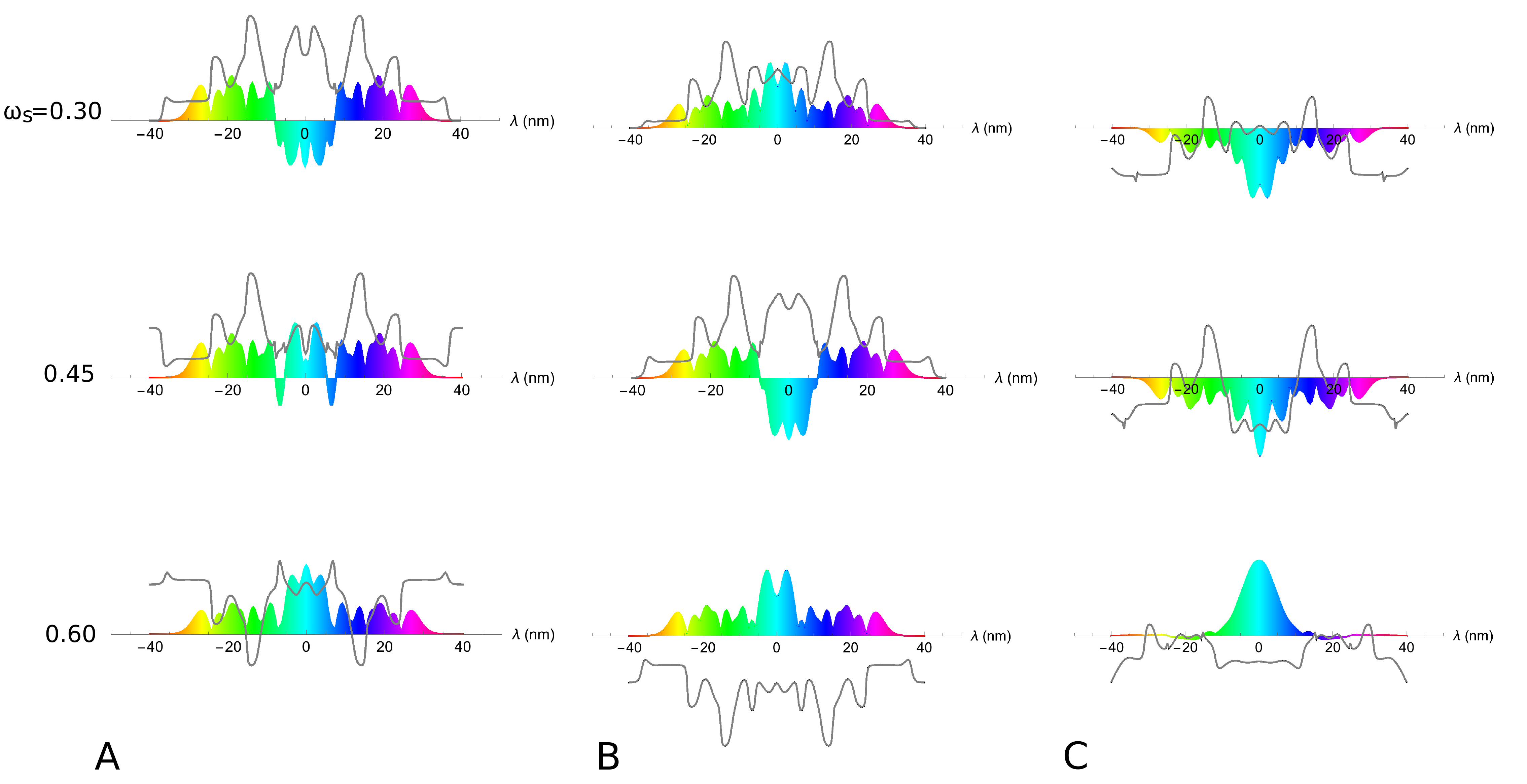}
  \caption{Shape of the LO which has to be set for measuring  the probe at three different frequencies $\omega_S$ in the case of a Gaussian pump spectrum for three networks. A,B and C stand for periodic, shortcuts and random network.}
  \label{fig:LO1}
\end{figure}

\begin{figure}
  \includegraphics[width=\linewidth]{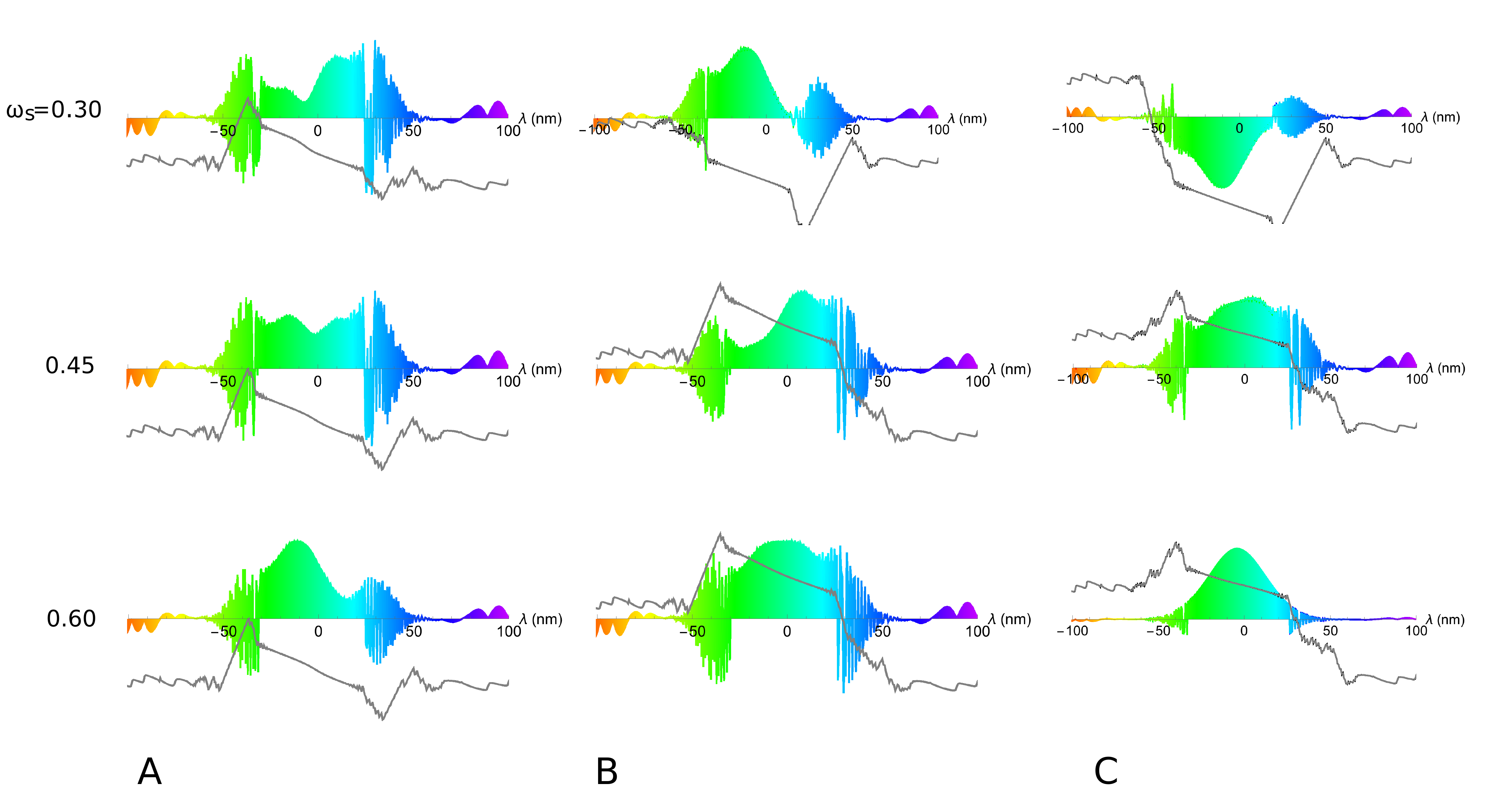}
  \caption{Shape of the LO which has to be set for measuring  the probe at three different frequencies $\omega_S$ when pump shaping with a $0.5$ mm crystal is considered. }
  \label{fig:LO}
\end{figure}

\begin{figure}
  \includegraphics[width=\linewidth]{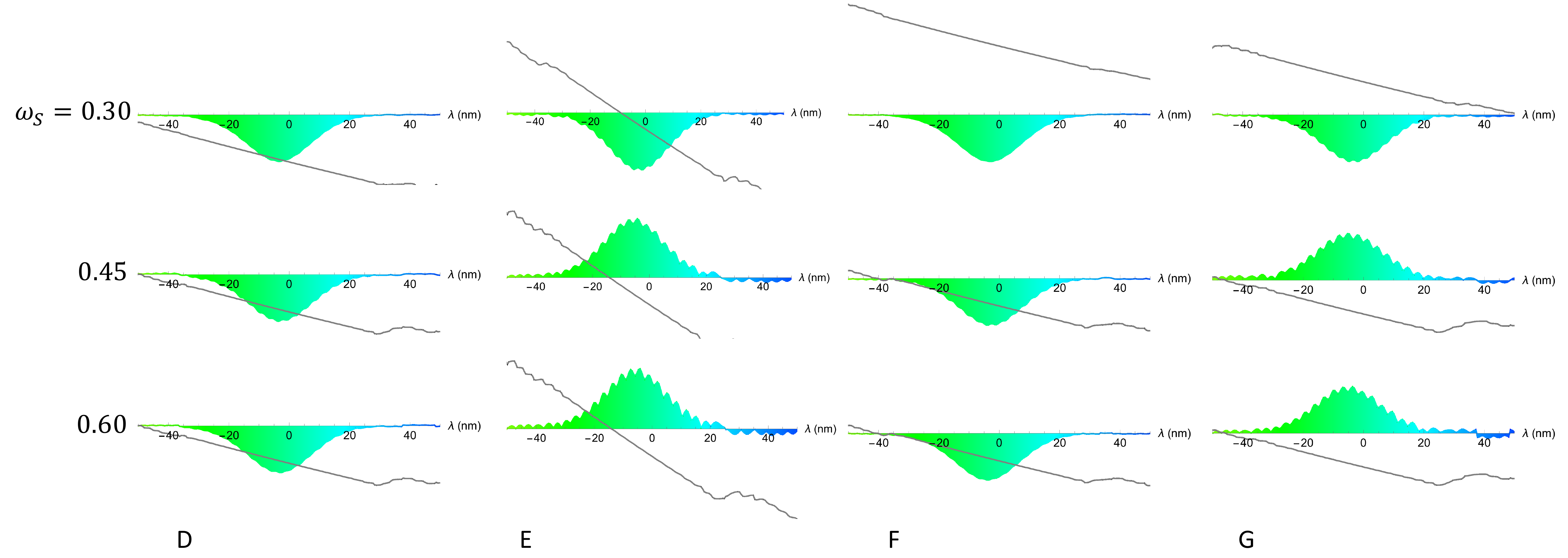}
  \caption{Shape of the LO which has to be set for measuring  the probe at three different frequencies $\omega_S$ when pump shaping with a $1.5$ mm crystal is considered for  Barab\'{a}si-Albert  (D), the Watts–Strogatz (E),  the Scannell (F)  and the Lisseau (G) networks. }
  \label{fig:LO2}
\end{figure}

We finally report, in Fig.\ref{fig:R}, the first 10 non-zero values of the N+1-th row of $R_{1f}$ at different frequencies  $\omega_S$. The N+1-th  row identifies the shape of the mode encoding the probe in term of the supermodes basis, the resulting shape has to be set as the shape of the LO in homodyne detection in order to measure  $\langle n(t)\rangle$. Examples of the LO shapes are shown in Fig. \ref{fig:LO1},  \ref{fig:LO} and \ref{fig:LO2} . As in the case of simulation of the networks dynamics, longer crystals allow for reduced LO bandwidth. In any case all the LO shapes calculated in the \ref{fig:LO1},  \ref{fig:LO} and \ref{fig:LO2}  are experimentally doable.

\section{Probing network from the entropy of the probe}
 An interesting concept, developed in \cite{galve2013} is whether a simple, resource friendly, measurement protocol can distinguish different classes of complex networks. It is indeed possible to distinguish among Random Regular, Erd\"os-Renyi and Scale-Free bosonic networks from features displayed by the averaged single node entropy $\langle S_h\rangle$, where the average is taken over all nodes of the same connectivity  $h$. Interestingly, also a set-up where only the entropy of a probe is measured allows to assess  such
different classes of networks \cite{galve2013}. In this set-up, the probe is sequentially attached to more and more network nodes (increasing its connectivity $h$) and measuring the probe's entropy as a function of $h$ suffices to discriminate network classes. 
The required von Neumann entropy is easy to access through the covariance matrix of a single node (or the probe) $\sigma=\{\{\alpha,\gamma\},\{\gamma,\beta\}\}$, and reads $S=(\mu+1/2)\log(\mu+1/2)-(\mu-1/2)\log(\mu-1/2)$ and $\mu=\sqrt{\alpha\beta-\gamma^2}$ \cite{agarwal}. The protocol is valid for a network in the ground state, but is robust for small temperatures, so the implementation devised above can be easily applied.

 \section{Conclusions and outlook}
In conclusion  we have shown a protocol for implementing reconfigurable experimental  quantum networks with a complex topology in a multimode quantum optical setup.  The temporal evolution of networks of various topology can be implemented by tailoring multimode squeezing operations and multimode measurements. This can be done via an optimization protocol on the pump shape for the parametric process and mode selective homodyne measurements.  Moreover the probing with one additional oscillator can be simulated: the recovered behaviours of the spectral density are able to univocally identify the corresponding network topology. Also probing based on entropy measurements is realizable. 
The proposed protocol establishes a connection between the field of continuous variables  experimental quantum optics and quantum complex network theory and it shows, for the first time, the full reconfiguratibility of an optical setup in implementing arbitrary B-M decompositions associated  with topologies which range from regular to complex.  \\
We have considered networks both from well-established theroretical models and from real biological and community examples.
In order to assess the complexity of the generated networks we are considering geometries involving at least of 40-50 nodes, this is for example the requirement for univocally identifying the reconstructed $J(\omega)$ with a non-trivial graphical structure. The experimental setup already demonstrated the capability of addressing networks made of 12 nodes \cite{Roslund14}.  
While the spectral shape of the required LO for the probing of $J(\omega)$ is within the experimental reach, the required shapes for following the network dynamics will require more experimental effort,  depending on the precision we want to reach for the simulated quantities.
The extension of involved modes numbers up to 50 is within the capability of the present setup by improving the pulse shaped homodyne detection procedure via the coherent broadening of LO spectrum  in order to address the squeezed high-order modes. Moreover the production of  multimode quantum states involving a large number of frequency- or temporal- modes   has already been demonstrated  in other experimental setups \cite{Yokoyama13,Chen14}. Hence the merging of techniques based on  spectral and temporal multiplexing can allow  the implementation of larger quantum complex networks.  \\ The quantum character of the networks is provided by the initialization of the nodes in quantum states or establishing entanglement correlations between different nodes. This condition already enables experimental study of control, engineering and probing of the network \cite{Moore16,Nokkala16,galve2013}, generation and protection of quantum resources \cite{Kostak07,Manzano13}, quantum synchronization \cite{Manzano13sy}, memory effects and dynamics of open quantum system  \cite{0034-4885-77-9-094001,RevModPhys.88.021002,RevModPhys.89.015001,Vasile14}. Also, as in the case of classical networks \cite{Gao2016}, the resilience of quantum task over network topology can be investigated. In particular quantum  communication tasks, like entanglement percolation \cite{Acin07,PhysRevLett.103.240503}, and their extension to multiparty protocols, like secret sharing,  \cite{Hillery99} can be studied. 
The subject of this work is the topological complexity but the present setup with a supplementary tool, i.e. mode-dependent coherent single photon subtractor, can be exploited in future works to demonstrate possible computational complexity, i.e. a form of quantum advantage. 
The mode-dependent coherent single photon subtraction, which has been recently experimentally demonstrated by some of the authors \cite{Ra17},  allows for the introduction of non-Gaussian processes in the network \cite{Walschaers17} and a specific circuit including multimode squeezing and single photon operations recently demonstrated to  be hard to sample \cite{Chabaud17}. The strategy can be also exploited to test proposed models of quantum enhanced transport \cite{Faccin14,Walschaers13} via the propagation of non-Gaussian excitation over complex graphs.\\
A second  supplementary tool which can be added  is multi-pixel  homodyne detector which enables simultaneous measurement of  several network components. Here after mixing the signal with the LO in the 50:50 BS  the spectral components of the light are spatially dispersed and focussed on two linear array of detectors.  Subtracting signals from the detectors corresponding to the same colours allows for simultaneous homodyne detection at each band. Pulse shaping on the LO also allows to independently choose the phase of the quadrature to be measured by independently modifying the phase of each spectral band. Finally  computer post-processing provides an additional basis change \cite{Ferrini13}.  
The simultaneous multimode analysis can be exploited for implementing a multi-probes scenario and moreover to test  measurement based continuous variables quantum information protocols over complex graphs.

\section*{Acknowledgements}
J.N., F.G., R.Z., S.M. and J.P. acknowledge financial support from the Horizon 2020 EU collaborative projects QuProCS (Grant Agreement No. 641277). F.G. and R.Z. acknowledge MINECO/AEI/FEDER through projects NoMaQ FIS2014-60343-P, QuStruct FIS2015-66860-P and EPheQuCS FIS2016-78010-P.  N.T. is  members of the Institut Universitaire de France. F.A., V.P., and N.T. acknowledge financial support from the European Union Grant QCUMbER (No. 665148).
V.P. acknowledges financial support from Emergence Sorbonne Universit\'{e} project PiCQuNet.
\section*{References}

\bibliography{complex_network-8}

\begin{thebibliography}{10}

\bibitem{RMPBarabasi}
R.~Albert and A.-L. Barab\'asi.
\newblock Statistical mechanics of complex networks.
\newblock {\em Rev. Mod. Phys.}, 74:47--97, Jan 2002.

\bibitem{Newmann}
M.~Newman, A.-L. Barab\'asi, and D.~J. Watts.
\newblock {\em The Structure and Dynamics of Networks}.
\newblock Princeton University Press, 2006.

\bibitem{Newman2}
M.~Newman.
\newblock {\em Networks, An Introduction}.
\newblock Oxford University Press, 2010.

\bibitem{Trabesinger}
A.~Trabesinger.
\newblock Complexity.
\newblock {\em Nat. Phys.}, 8:13--13, Jan 2012.
\newblock and following articles in this insight issue.

\bibitem{Goh02}
K.-I. Goh, E.~Oh, H.~Jeong, B.~Kahng, and D.~Kim.
\newblock Classification of scale-free networks.
\newblock {\em Proceedings of the National Academy of Sciences},
  99(20):12583--12588, 2002.

\bibitem{Amaral00}
L.~A.~N. Amaral, A.~Scala, M.~Barthélémy, and H.~E. Stanley.
\newblock Classes of small-world networks.
\newblock {\em Proceedings of the National Academy of Sciences},
  97(21):11149--11152, 2000.

\bibitem{Rosvall08}
M.~Rosvall and C.~T. Bergstrom.
\newblock Maps of random walks on complex networks reveal community structure.
\newblock {\em Proceedings of the National Academy of Sciences},
  105(4):1118--1123, 2008.

\bibitem{Bianconi2015}
G.~Bianconi.
\newblock Interdisciplinary and physics challenges of network theory.
\newblock {\em EPL (Europhysics Letters)}, 111:56001, 2015.

\bibitem{Biamonte17}
J.~Biamonte, M.~Faccin, and M.~De~Domenico.
\newblock Complex networks: from classical to quantum.
\newblock {\em arXiv preprint arXiv:1702.08459}, 2017.

\bibitem{Acin07}
A.~Acin, J.~I. Cirac, and M.~Lewenstein.
\newblock Entanglement percolation in quantum networks.
\newblock {\em Nat. Phys.}, 3(4):256--259, Apr 2007.

\bibitem{Faccin14}
M.~Faccin, P.~Migda\l{}, T.~H. Johnson, V.~Bergholm, and J.~D. Biamonte.
\newblock Community detection in quantum complex networks.
\newblock {\em Phys. Rev. X}, 4:041012, Oct 2014.

\bibitem{Paparo13}
G.~D. Paparo, M.~Muller, F.~Comellas, and M.~A. Martin-Delgado.
\newblock Quantum google in a complex network.
\newblock {\em Sci. Rep.}, 3:2773, 2013.

\bibitem{Perseguers10}
S.~Perseguers, M.~Lewenstein, A.~Acin, and J.~I. Cirac.
\newblock Quantum random networks.
\newblock {\em Nat. Phys.}, 6(7):539--543, Jul 2010.

\bibitem{PhysRevLett.101.175702}
L.~Jahnke, J.~W. Kantelhardt, R.~Berkovits, and S.~Havlin.
\newblock Wave localization in complex networks with high clustering.
\newblock {\em Phys. Rev. Lett.}, 101:175702, Oct 2008.

\bibitem{PhysRevE.87.022104}
A.~Halu, S.~Garnerone, A.~Vezzani, and G.~Bianconi.
\newblock Phase transition of light on complex quantum networks.
\newblock {\em Phys. Rev. E}, 87:022104, Feb 2013.

\bibitem{0953-4075-34-23-314}
R.~Burioni, D.~Cassi, M.~Rasetti, P.~Sodano, and A.~Vezzani.
\newblock Bose-einstein condensation on inhomogeneous complex networks.
\newblock {\em J. Phys. B At. Mol. Opt. Phys.}, 34(23):4697, 2001.

\bibitem{MULKEN201137}
O.~Mulken and A.~Blumen.
\newblock Continuous-time quantum walks: Model for coherent transport on
  complex networks.
\newblock {\em Phys. Rep.}, 502:37--87, 2011.

\bibitem{Kimble08}
H.~J. Kimble.
\newblock The quantum internet.
\newblock {\em Nature}, 453:1023--1030, 2008.

\bibitem{PhysRevLett.103.240503}
M.~Cuquet and J.~Calsamiglia.
\newblock Entanglement percolation in quantum complex networks.
\newblock {\em Phys. Rev. Lett.}, 103:240503, Dec 2009.

\bibitem{Rossi14}
M.~Rossi, D.~Bru{\ss}, and C.~Macchiavello.
\newblock Hypergraph states in grover's quantum search algorithm.
\newblock {\em Physica Scripta}, 2014(T160):014036, 2014.

\bibitem{Bianconi16}
G.~Bianconi and C.~Rahmede.
\newblock Network geometry with flavor: From complexity to quantum geometry.
\newblock {\em Phys. Rev. E}, 93:032315, Mar 2016.

\bibitem{Kato14}
Y.~Kato and N.~Yamamoto.
\newblock Structure identification and state initialization of spin networks
  with limited access.
\newblock {\em New J. Phys.}, 16(2):023024, 2014.

\bibitem{Jurcevic14}
P.~Jurcevic, P.~Lanyon, B. P.and~Hauke, C.~Hempel, P.~Zoller, R.~Blatt, and
  C.~F. Roos.
\newblock Quasiparticle engineering and entanglement propagation in a quantum
  many-body system.
\newblock {\em Nature}, 511(7508):202--205, Jul 2014.

\bibitem{Elliott16}
T.~J. Elliott and I.~B. Mekhov.
\newblock Engineering many-body dynamics with quantum light potentials and
  measurements.
\newblock {\em Phys. Rev. A}, 94:013614, Jul 2016.

\bibitem{Moore16}
D.~W. Moore, T.~Tufarelli, M.~Paternostro, and A.~Ferraro.
\newblock Quantum state reconstruction of an oscillator network in an
  optomechanical setting.
\newblock {\em Phys. Rev. A}, 94:053811, Nov 2016.

\bibitem{Gokler17}
C.~Gokler, S.~Lloyd, P.~Shor, and K.~Thompson.
\newblock Efficiently controllable graphs.
\newblock {\em Phys. Rev. Lett.}, 118:260501, Jun 2017.

\bibitem{Yung05}
M.-H. Yung and S.~Bose.
\newblock Perfect state transfer, effective gates, and entanglement generation
  in engineered bosonic and fermionic networks.
\newblock {\em Phys. Rev. A}, 71:032310, Mar 2005.

\bibitem{Kostak07}
V.~Kostak, G.~M. Nikolopoulos, and I.~Jex.
\newblock Perfect state transfer in networks of arbitrary topology and coupling
  configuration.
\newblock {\em Phys. Rev. A}, 75:042319, Apr 2007.

\bibitem{Manzano13}
G.~Manzano, F.~Galve, and R.~Zambrini.
\newblock Avoiding dissipation in a system of three quantum harmonic
  oscillators.
\newblock {\em Phys. Rev. A}, 87:032114, Mar 2013.

\bibitem{Arena08}
A.~Arenas, A.~Díaz-Guilera, J.~Kurths, Y.~Moreno, and C.~Zhou.
\newblock Synchronization in complex networks.
\newblock {\em Physics Reports}, 469(3):93 -- 153, 2008.

\bibitem{Dofler13}
F.~Dorfler, M.~Chertkov, and F.~Bullo.
\newblock Synchronization in complex oscillator networks and smart grids.
\newblock {\em Proceedings of the National Academy of Sciences},
  110(6):2005--2010, 2013.

\bibitem{Manzano13sy}
G.~Manzano, F.~Galve, G.~L. Giorgi, E.~Hern\'{a}ndez-Garc\'{i}a, and
  R.~Zambrini.
\newblock Synchronization, quantum correlations and entanglement in oscillator
  networks.
\newblock {\em Sci. Rep.}, 3:1439, 2013.

\bibitem{Bellomo17}
B.~Bellomo, G.~L. Giorgi, G.~M. Palma, and R.~Zambrini.
\newblock Quantum synchronization as a local signature of super- and
  subradiance.
\newblock {\em Phys. Rev. A}, 95:043807, Apr 2017.

\bibitem{Nokkala16}
J.~Nokkala, F.~Galve, R.~Zambrini, S.~Maniscalco, and J.~Piilo.
\newblock Complex quantum networks as structured environments: engineering and
  probing.
\newblock {\em Sci. Rep.}, 6:26861, May 2016.

\bibitem{Roslund14}
J.~Roslund, R.~Medeiros~de Ara\'ujo, S.~Jiang, C.~Fabre, and N.~Treps.
\newblock Wavelength-multiplexed quantum networks with ultrafast frequency
  combs.
\newblock {\em Nat. Photon.}, 8:109, 2014.

\bibitem{Sameti17}
M.~Sameti, A.~Poto\ifmmode~\check{c}\else \v{c}\fi{}nik, D.~E. Browne,
  A.~Wallraff, and M.~J. Hartmann.
\newblock Superconducting quantum simulator for topological order and the toric
  code.
\newblock {\em Phys. Rev. A}, 95:042330, Apr 2017.

\bibitem{Robens15}
C.~Robens, W.~Alt, D.~Meschede, C.~Emary, and A.~Alberti.
\newblock Ideal negative measurements in quantum walks disprove theories based
  on classical trajectories.
\newblock {\em Phys. Rev. X}, 5:011003, Jan 2015.

\bibitem{Qiang16}
X.~Qiang, T.~Loke, A.~Montanaro, K.~Aungskunsiri, X.~Zhou, J.~L. O'Brien, J.~B.
  Wang, and J.~C.~F. Matthews.
\newblock Efficient quantum walk on a quantum processor.
\newblock {\em Nature Communications}, 7:11511 EP --, May 2016.

\bibitem{Gerke15}
S.~Gerke, J.~Sperling, W.~Vogel, Y.~Cai, J.~Roslund, N.~Treps, and C.~Fabre.
\newblock Full multipartite entanglement of frequency-comb gaussian states.
\newblock {\em Phys. Rev. Lett.}, 114:050501, Feb 2015.

\bibitem{Cai17}
Y.~Cai, J.~Roslund, G.~Ferrini, F.~Arzani, X.~Xu, C.~Fabre, and N.~Treps.
\newblock Multimode entanglement in reconfigurable graph states using optical
  frequency combs.
\newblock {\em Nat. Commun.}, 8:15645, June 2017.

\bibitem{galve2013}
A.~Cardillo, F.~Galve, D.~Zueco, and J.~G\'omez-Garde\~nes.
\newblock Information sharing in quantum complex networks.
\newblock {\em Phys. Rev. A}, 87:052312, May 2013.

\bibitem{0034-4885-77-9-094001}
A.~Rivas, S.~F. Huelga, and M.~B. Plenio.
\newblock Quantum non-markovianity: characterization, quantification and
  detection.
\newblock {\em Rep. Prog. Phys.}, 77(9):094001, 2014.

\bibitem{RevModPhys.88.021002}
H.-P. Breuer, E.-M. Laine, J.~Piilo, and B.~Vacchini.
\newblock Colloquium.
\newblock {\em Rev. Mod. Phys.}, 88:021002, Apr 2016.

\bibitem{RevModPhys.89.015001}
I.~de~Vega and D.~Alonso.
\newblock Dynamics of non-markovian open quantum systems.
\newblock {\em Rev. Mod. Phys.}, 89:015001, Jan 2017.

\bibitem{Liu:2011aa}
B.-H. Liu, L.~Li, Y.-F. Huang, C.-F. Li, G.-C. Guo, E.-M. Laine, H.-P. Breuer,
  and J.~Piilo.
\newblock Experimental control of the transition from markovian to
  non-markovian dynamics of open quantum systems.
\newblock {\em Nat. Phys.}, 7(12):931--934, 12 2011.

\bibitem{Vasile14}
R.~Vasile, F.~Galve, and R.~Zambrini.
\newblock Spectral origin of non-markovian open-system dynamics: A finite
  harmonic model without approximations.
\newblock {\em Phys. Rev. A}, 89:022109, Feb 2014.

\bibitem{Paparo12}
G.~D. Paparo and M.~A. Martin-Delgado.
\newblock Google in a quantum network.
\newblock {\em Sci. Rep.}, 2:444, 2012.

\bibitem{Walschaers16rev}
M.~Walschaers, F.~Schlawin, T.~Wellens, and A.~Buchleitner.
\newblock Quantum transport on disordered and noisy networks: An interplay of
  structural complexity and uncertainty.
\newblock {\em Annu. Rev. Conden. Ma. P.}, 7(1):223--248, 2016.

\bibitem{Braunstein05}
S.~L. Braunstein.
\newblock Squeezing as an irreducible resource.
\newblock {\em Phys. Rev. A}, 71:055801, May 2005.

\bibitem{DeGosson06}
M.~A. De~Gosson.
\newblock {\em Symplectic Geometry and Quantum Mechanics}.
\newblock Birkhäuser, 2006.
\newblock Proposition 2.13 and its proof.

\bibitem{Medeiros14}
R.~Medeiros~de Ara\'ujo, J.~Roslund, Y.~Cai, G.~Ferrini, C.~Fabre, and
  N.~Treps.
\newblock Full characterization of a highly multimode entangled state embedded
  in an optical frequency comb using pulse shaping.
\newblock {\em Phys. Rev. A}, 89:053828, May 2014.

\bibitem{Arzani17}
F.~Arzani, C.~Fabre, and N.~Treps.
\newblock Versatile engineering of multimode squeezed states by optimizing the
  pump spectral profile in spontaneous parametric down-conversion.
\newblock {\em Phys. Rev. A}, 97:033808, Mar 2018.

\bibitem{Barabasi99}
A.-L. Barab{\'a}si and R.~Albert.
\newblock Emergence of scaling in random networks.
\newblock {\em Science}, 286(5439):509--512, Sep 1999.

\bibitem{Watts98}
D.~J. Watts and S.~H. Strogatz.
\newblock Collective dynamics of 'small-world' networks.
\newblock {\em Nature}, 393(6684):440--442, June 1998.

\bibitem{Scannell99}
J.~Scannell, G.~Burns, C.~Hilgetag, M.~O'Neil, and M.~Young.
\newblock The connectional organization of the cortico-thalamic system of the
  cat.
\newblock {\em Cerebral Cortex}, 9(3):277--299, 1999.

\bibitem{Lusseau03}
D.~Lusseau, K.~Schneider, O.~J. Boisseau, P.~Haase, E.~Slooten, and S.~M.
  Dawson.
\newblock The bottlenose dolphin community of doubtful sound features a large
  proportion of long-lasting associations.
\newblock {\em Behavioral Ecology and Sociobiology}, 54(4):396--405, Sep 2003.

\bibitem{Davies76}
E.~Davies.
\newblock {\em Quantum Theory of Open Systems}.
\newblock Academic Press, 1976.

\bibitem{Breuer07}
H.-P. Breuer and F.~Petruccione.
\newblock {\em The Theory of Open Quantum Systems}.
\newblock Oxford University Press, 2007.

\bibitem{agarwal}
G.~S. Agarwal.
\newblock Entropy, the wigner distribution function, and the approach to
  equilibrium of a system of coupled harmonic oscillators.
\newblock {\em Phys. Rev. A}, 3:828--831, Feb 1971.

\bibitem{Yokoyama13}
S.~Yokoyama, R.~Ukai, S.~C. Armstrong, C.~Sornphiphatphong, T.~Kaji, S.~Suzuki,
  J.-i. Yoshikawa, H.~Yonezawa, N.~C. Menicucci, and A.~Furusawa.
\newblock Ultra-large-scale continuous-variable cluster states multiplexed in
  the time domain.
\newblock {\em Nat. Photon.}, 7:982, Dec 2013.

\bibitem{Chen14}
M.~Chen, N.~C. Menicucci, and O.~Pfister.
\newblock Experimental realization of multipartite entanglement of 60 modes of
  a quantum optical frequency comb.
\newblock {\em Phys. Rev. Lett.}, 112:120505, Mar 2014.

\bibitem{Gao2016}
J.~Gao, B.~Barzel, and A.-L. Barab{\'a}si.
\newblock Universal resilience patterns in complex networks.
\newblock {\em Nature}, 530:307 EP --, Feb 2016.

\bibitem{Hillery99}
M.~Hillery, V.~Bu\ifmmode~\check{z}\else \v{z}\fi{}ek, and A.~Berthiaume.
\newblock Quantum secret sharing.
\newblock {\em Phys. Rev. A}, 59:1829--1834, Mar 1999.

\bibitem{Ra17}
Y.-S. Ra, C.~Jacquard, A.~Dufour, C.~Fabre, and N.~Treps.
\newblock Tomography of a mode-tunable coherent single-photon subtractor.
\newblock {\em Phys. Rev. X}, 7:031012, Jul 2017.

\bibitem{Walschaers17}
M.~Walschaers, C.~Fabre, V.~Parigi, and N.~Treps.
\newblock Entanglement and wigner function negativity of multimode non-gaussian
  states.
\newblock {\em Phys. Rev. Lett.}, 119:183601, Oct 2017.

\bibitem{Chabaud17}
U.~Chabaud, T.~Douce, D.~Markham, P.~Van~Loock, E.~Kashefi, and G.~Ferrini.
\newblock Continuous-variable sampling from photon-added or photon-subtracted
  squeezed states.
\newblock {\em arXiv preprint arXiv:1707.09245}, 2017.

\bibitem{Walschaers13}
M.~Walschaers, J.~F.-d.-C. Diaz, R.~Mulet, and A.~Buchleitner.
\newblock Optimally designed quantum transport across disordered networks.
\newblock {\em Phys. Rev. Lett.}, 111:180601, Oct 2013.

\bibitem{Ferrini13}
G.~Ferrini, J.~P. Gazeau, T.~Coudreau, C.~Fabre, and N.~Treps.
\newblock Compact gaussian quantum computation by multi-pixel homodyne
  detection.
\newblock {\em New J. Phys.}, 15:093015, 2013.

\end{thebibliography}

\bibliographystyle{unsrt-abbrv}

\end{document}